\documentclass[12pt,english,a4paper]{article}
\pdfoutput=1

\usepackage{mathtools,amsthm,amssymb,babel,cite,tabu,subcaption,tikz,heptit}

\usepackage[hidelinks,pagebackref=true]{hyperref}
\usepackage{cleveref}

\usetikzlibrary{decorations.markings}
\tikzset{->-/.style={decoration={markings,mark=at position .5 with {\arrow{>}}},postaction={decorate}}}

\newcommand\be{\begin{equation}\begin{aligned}}
\newcommand\ee{\end{aligned}\end{equation}}

\newcommand\hg{\hat g}
\newcommand\gv{\mathrm{GV}}
\newcommand\hgv{\widehat \gv_{\hg,\beta}}
\newcommand\gvp{\gv'^+_{\hg,\beta}}
\newcommand\gvm{\gv'^-_{\hg,\beta}}

\newcommand{\eu}[1]{\operatorname{e}\left( #1 \right)}
\DeclareMathOperator\tr{Tr}

\newcommand\hil{\mathcal H}
\newcommand\ud{\mathrm d}
\newcommand\iu{\mathrm i}

\newcommand\SU{\mathrm {SU}}
\newcommand\SO{\mathrm {SO}}
\newcommand\U{\mathrm U}
\newcommand\FF{\mathcal F}
\newcommand\NN{\mathcal N}

\newcommand\mathown{\mathbb}

\newcommand\IX{\mathown X}
\newcommand\CP{\mathown {CP}}
\newcommand\RP{\mathown {RP}}
\newcommand\IR{\mathown R}
\newcommand\IC{\mathown C}
\newcommand\IZ{\mathown Z}
\newcommand\IS{\mathown S}

\preprint{IFT-UAM/CSIC-15-055\\SISSA 25/2015 MATE}

\title{Towards a gauge theory interpretation\\of the real topological string}

\author[*]{Hirotaka Hayashi}
\author[**]{Nicol\`o Piazzalunga}
\author[*]{Angel M.~Uranga}

\affil[*]{Instituto de F\'{\i}sica Te\'orica IFT-UAM/CSIC\affilcr
C/ Nicol\'as Cabrera 13-15, Universidad Aut\'onoma de Madrid, 28049 Madrid, Spain}
\affil[**]{International School for Advanced Studies (SISSA)\affilcr
and Istituto Nazionale di Fisica Nucleare (INFN) - Sezione di Trieste\affilcr
Via Bonomea 265, 34136 Trieste, Italy}

\abstract{
We consider the real topological string on certain non-compact toric Calabi-Yau three-folds $\IX$, in its physical realization describing an orientifold of type IIA on $\IX$ with an O4-plane and a single D4-brane stuck on top. The orientifold can be regarded as a new kind of surface operator on the gauge theory with 8 supercharges arising from the singular geometry. We use the M-theory lift of this system to compute the real Gopakumar-Vafa invariants (describing  wrapped M2-brane BPS states) for diverse geometries. We show that the real topological string amplitudes pick up certain signs across flop transitions, in a well-defined pattern consistent with continuity of the real BPS invariants. We further give some preliminary proposals of an intrinsically gauge theoretical description of the effect  of the surface operator in the gauge theory partition function.
}

\begin{document}

\maketitle

\tableofcontents

\section{Introduction}

A most interesting connection between gauge theory and string theory is the relation
between non-perturbative instanton corrections in 4d/5d gauge theories with 8 supercharges\cite{Nekrasov:2002qd, Nekrasov:2003rj},
and the topological string partition functions on non-compact toric Calabi-Yau (CY) three-folds $\IX$
(in their formulation in terms of BPS invariants\cite{Gopakumar:1998ii,Gopakumar:1998jq,Ooguri:1999bv}).
Basically, the BPS M2-branes on compact 2-cycles of $\IX$ are instantons of the gauge theory arising from M-theory on $\IX$.

In this paper we are interested in extending this correspondence to systems with orientifold projections.
A natural starting point is the real topological string,
introduced in \cite{Walcher:2007qp} and studied in compact examples in \cite{Krefl:2008sj}
and in non-compact CY three-folds in \cite{Krefl:2009md,Krefl:2009mw}.
This real topological string is physically related to type IIA on a CY three-fold $\IX$ quotiented by an orientifold,
given by an antiholomorphic involution $\sigma$ on $\IX$ and a flip on two of the 4d spacetime coordinates (say $x^2$, $x^3$);
it thus introduces an O4-plane, spanning the lagrangian 3-cycle given by the fixed point set,\footnote{In this paper we focus on cases with non-trivial fixed point sets; freely acting orientifolds could be studied using similar ideas.}
and the two fixed 4d spacetime dimensions $x^0$, $x^1$.
In addition, the topological tadpole cancellation \cite{Walcher:2007qp} requires the introduction of a single stuck D4-brane
on top of the (negatively charged) O4-plane, producing local cancellation of RR charge.

The M-theory lift of this type IIA configuration \cite{unoriented-gv} corresponds to
a freely acting quotient of M-theory on $\IX \times \IS^1$,
in which the action $\sigma$ on $\IX$ (and the flip of two 4d coordinates) is accompanied by a half-period shift along the $\IS^1$.
This M-theory lift provides a reinterpretation of the real topological string partition function in terms of real BPS invariants,
which are essentially given by a combination of the parent Gopakumar-Vafa (GV) BPS invariants,
weighted by the $\pm 1$ eigenvalue of the corresponding state under the orientifold action.

This orientifolding can be applied in the context of type IIA/M-theory on non-compact toric CY three-folds $\IX$ which realizes 5d gauge theories. More specifically, one should consider the 5d gauge theory compactified to 4d, with an orientifold acting non-trivially as a shift on the $\IS^1$. Since the orientifold plane is real codimension 2 in the 4d Minkowski dimensions, the system describes the gauge theory in the presence of a surface defect. Certain surface operators have been studied in \cite{Gukov:2006jk, Gukov:2007ck,Gukov:2008sn, Gaiotto:2009fs, Kanno:2011fw}, also in the context of M-theory/gauge theory correspondence, describing them by the introduction of D2-branes/M2-branes in the brane setup or D4-branes in the geometric engineering \cite{Ooguri:1999bv, Dimofte:2010tz, Taki:2010bj}. An important difference with our discussion is that the holonomy of our surface operators is an outer automorphism of the original gauge group.
In our case, the properties of the gauge theory in this orientifold background are implicitly defined by the real topological vertex, even though they should admit an eventual intrinsic gauge-theoretical description. Hence, one can regard our real topological string results as a first step in the study of a novel kind of gauge theory surface operators.

Our strategy is as follows:
\begin{itemize}
\item For concreteness, we focus on the geometry which realizes a pure $\SU(N)$ gauge theory, and other similarly explicit examples. We construct the real topological string on non-compact toric CY three-folds $\IX$ by using the real topological vertex formalism \cite{Krefl:2009mw}. 

\item In the M-theory interpretation, the real topological string amplitudes correspond to a one-loop diagram of a set of 5d BPS particles from wrapped M2-branes, suitably twisted by the orientifold action as they propagate on the $\IS^1$. An important point is that the effect of the orientifold action arises only after the compactification to 4d on $\IS^1$, so the 5d picture is identical to the parent theory. Therefore, the correspondence between M-theory and 5d $\NN=1$ gauge theory is untouched. 

\item The corresponding statement on the gauge theory side is that the 4d partition function of the gauge theory in the presence of the orientifold surface defect must be given by the compactification of the original 5d gauge theory on $\IS^1$, but with modified periodic/anti-periodic boundary conditions for fields which are even/odd under the orientifold action. This can in principle be implemented as the computation of the Witten index with an extra twist operator in the trace. This kind of operator has not appeared in the literature. We use the comparison of the oriented and real topological vertex partition functions to better understand the nature and action of this operator on the gauge theory. 
\end{itemize}

Even though we do not achieve a completely successful gauge theoretical definition of the orientifold operation,  we obtain a fairly precise picture of this action in some concrete situations. Moreover, our discussion of the topological vertex amplitudes reveals new properties in the unoriented case. 

\medskip

The rest of the paper is organized as follows. 
In \cref{sec:review-oriented} we review the M-theory/gauge theory correspondence in the oriented case:
in \cref{sec:top-oriented} we review the topological vertex computation of topological string partition functions
on local CY three-folds,
in \cref{sec:gauge-oriented} we describe the computation of the gauge theoretic Nekrasov partition function via localization.
In \cref{sec:unoriented} we review the computation of real topological string amplitudes:
\cref{sec:gen-unoriented} introduces some general considerations of unoriented theories,
and \cref{sec:top-unoriented} describes the real topological string computation using the real topological vertex.
Explicit examples are worked out in \cref{sec:examples-unoriented},
like the conifold ($\U(1)$ gauge theory) in \cref{sec:coni-unoriented}, where we correct some typos in the previously known result,
and the pure $\SU(N)$ theories in \cref{sec:sun-unoriented},
where we also discuss their behavior under flop transitions.
In \cref{sec:towards-gauge} we describe a twisted Nekrasov partition function, whose structure is motivated by the action of the orientifold, and compare it with the real topological string partition function.
\Cref{conclusions} offers our conclusions.
\Cref{app:background} reviews aspects of the real topological string and the topological vertex formulation,
\cref{sec:enum} presents some new enumerative checks of the BPS integrality,
and \cref{sec:help} gathers some useful identities.

\section{Review of the oriented case}
\label{sec:review-oriented}

In this section we review the correspondence between BPS M2-brane invariants of M-theory
on a toric CY three-fold singularity $\IX$ and the supersymmetric gauge theory Nekrasov partition function \cite
{Iqbal:2003ix,Iqbal:2003zz,Eguchi:2003sj,Hollowood:2003cv,zhou-curve,Taki:2007dh}.
We will take advantage to introduce useful tools and notations to be used in the discussion of the orientifolded case.

\subsection{The topological string}
\label{sec:top-oriented}

\subsubsection{BPS expansion of the topological string}
\label{bps-top}

We start with a brief review of closed oriented topological string interpreted in terms of BPS states in M-theory \cite
{Gopakumar:1998ii,Gopakumar:1998jq}.
Consider type IIA on a CY three-fold $\IX$, which provides a physical realization of the topological A-model on $\IX$.
The genus $g$ topological string amplitude $F_g(t_i)$, which depends on the complexified K\"ahler moduli $t_i = a_i + \iu v_i$ with $a_i$ coming from the B-field and $v_i$ being the volume, computes the F-term\cite{Antoniadis:1993ze}
\be
\label{R2}
\int \ud^4 x \int \ud^4\theta \, F_g(t_i) \left(\mathcal W^2\right)^g \to \int \ud^4 x \, F_g(t_i) F_+^{2g-2} R_+^2,
\ee
where the second expression applies for $g>1$ only, and the $\NN=2$ Weyl multiplet is schematically
$\mathcal W= F_+ + \theta^2 R_+ + \cdots$, in terms of the self-dual graviphoton and curvature.
If we turn on a constant self-dual graviphoton background in the four non-compact dimensions
\be
\label{sdgrav}
F_+ = \frac\epsilon2 \ud x^1 \wedge \ud x^2 + \frac\epsilon2 \ud x^3 \wedge \ud x^4,
\ee
the sum may be regarded as the total A-model free-energy, with coupling $\epsilon$
\be
\FF (t_i) = \sum_{g=0}^\infty \epsilon^{2g-2} F_g(t_i).
\ee
This same quantity can be directly computed as a one-loop diagram of 5d BPS states in M-theory compactified on $\IX\times \IS^1$,
corresponding to 11d graviton multiplets and to M2-branes wrapped on holomorphic 2-cycles.
These states couple to the graviphoton background via their quantum numbers
under the $\SU(2)_L$ in the 5d little group $\SU(2)_L\times \SU(2)_R$.
Denoting by $\gv_{g,\beta}$ the multiplicity of BPS states corresponding to M2-branes
on a genus $g$ curve in the homology class $\beta$, we have an expression\footnote
{A subtlety regarding the reality condition of the $\epsilon$-background has been discussed in \cite{Dedushenko:2014nya}.}
\be
\label{gv-closed-d}
\FF = \sum_{\beta\in H_2 (\IX; \IZ)} \sum_{g=0}^\infty \sum_{m=1}^\infty \gv_{g,\beta} \frac1m
\left( 2 \iu \sinh \frac {m\epsilon}2 \right)^{2g-2} e^{\iu m \beta\cdot t},
\ee 
with $|e^{\iu m \beta\cdot t}| < 1$ so that the BPS state counting is well-defined. Namely, the computation is carried out in the large volume point in the K\"ahler moduli space. From the M-theory point of view, the real part of $t$ may be provided by the Wilson line along $\IS^1$, originated from the three-form $C_3$. Finally, the topological string partition function is defined as
\be
\label{top.partition}
Z_\text{top}=\exp \FF.
\ee

\subsubsection{Topological vertex formalism}
\label{sec:top-vert}

From now on we specialize to a particular class of M-theory background geometries, directly related to supersymmetric gauge theories.
M-theory on a CY three-fold singularity, in the decoupling limit, implements a geometric engineering realization of 5d gauge theory
with 8 supercharges, with gauge group and matter content determined by the singularity structure \cite{Morrison:1996xf, Intriligator:1997pq}.
Upon compactification on $\IS^1$, it reproduces type IIA geometric engineering\cite{Katz:1996fh, Katz:1997eq}.
This setup is therefore well-suited for the matching of gauge theory results
in terms of topological string amplitudes in the GV interpretation.

We will use the the best known tool at large volume point in moduli space
to compute topological string partition functions on local toric CY three-folds,
namely the topological vertex formalism\cite{Iqbal:2002we, Aganagic:2003db,Awata:2005fa,Iqbal:2007ii}.
This can be even used to define a refined version of the topological string amplitude \cite{Awata:2005fa,Iqbal:2007ii,Gukov:2007tf,Aganagic:2011mi,Aganagic:2012au,Choi:2012jz,Antoniadis:2013bja},
associated to the theory in a non self-dual graviphoton background
\be
F =\frac12 \epsilon_1 \ud x^1 \wedge \ud x^2 - \frac12 \epsilon_2 \ud x^3 \wedge \ud x^4.
\ee
The unrefined topological string amplitude is recovered for $\epsilon_1=-\epsilon_2$.
We will describe the refined topological vertex computation, but will eventually restrict to the unrefined case,
since only this is known in the unoriented case.
Happily, this suffices to illustrate our main points.

The basic idea is to regard the web diagram of the resolved singularity (the dual of the toric fan) as a Feynman diagram,
with rules to produce the topological string partition function.
Roughly, one sums over edges that correspond to Young diagrams $R_i$,
with propagators $\left(-e^{\iu t_i}\right)^{|R_i|}$ depending on K\"ahler parameters $t_i$,
and vertex functions expressed in terms of skew-Schur functions.
The formalism is derived from open-closed string duality and 3d Chern-Simons theory\cite
{Iqbal:2002we, Aganagic:2003db, Aganagic:2012hs}.

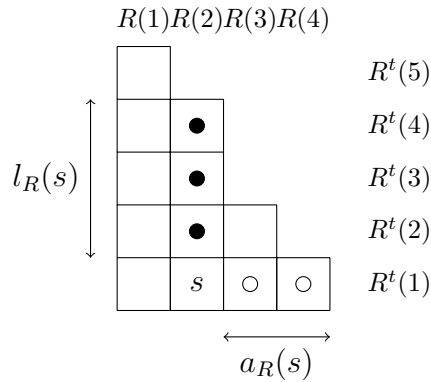
\begin{figure}[htb]
\centering

\begin{tikzpicture}[scale=.7]

\draw (0,0) grid (1,5);
\draw (1,0) grid (2,4);
\draw (2,0) grid (3,2);
\draw (3,0) grid (4,1);

\node at (1/2+1,1/2) {$s$};
\foreach \x in {2,...,3} \draw (\x+1/2,1/2) circle (4pt);
\foreach \y in {1,...,3} \filldraw (1/2+1,\y+1/2) circle (4pt);

\draw [<->] (-1/2,1) -- (-1/2,4);
\node[anchor=east] at (-1/2,2.5) {$l_R(s)$};

\draw [<->] (2,-1/2) -- (4,-1/2);
\node[anchor=north] at (3,-1/2) {$a_R(s)$};

\foreach \y in {1,...,5} \node[anchor=west] at (4+1/2,\y-1/2) {\footnotesize $R^t(\y)$};
\foreach \x in {1,...,4} \node at (\x-1/2,5+1/2) {\footnotesize $R(\x)$};
\end{tikzpicture}

\caption{Notation for arm and leg lengths of the box $s=(2,1)\in R$.}
\label{hook}
\end{figure}

Let us begin by introducing some useful definitions.
A Young diagram $R$ is defined by the numbers of boxes $R(i)$ in the $i^\text{th}$ column,
ordered as $R(1) \geq R(2) \geq \cdots \geq R(d) \geq R(d+1)=0$, see \cref{hook}.
We denote by $|R|=\sum_{i=1}^d R(i)$ the total number of boxes,
and by $\varnothing$ the empty diagram.
For a box $s =(i,j)\in R$, we also define
\be
a_R(i,j) := R^t(j)-i, \qquad l_R(i,j) := R(i)-j,
\ee
where $R^t$ denotes transpose, see \cref{hook}.

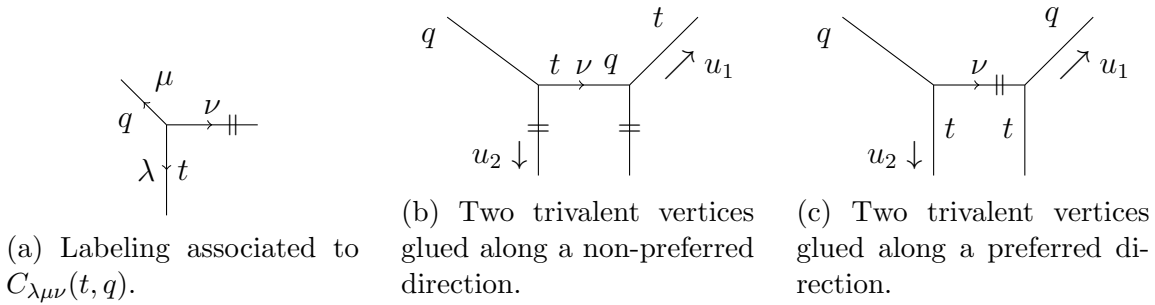
\begin{figure}[htb]
\centering

\begin{subfigure}[b]{.3\textwidth}
\centering
\begin{tikzpicture}[node distance=2cm, auto,scale=.3]
\draw[->-] (0,0) to (4,0);
\draw[->-] (0,0) to (-2,2);
\draw[->-] (0,0) to (0,-4);

\node[rotate=90] at (3,0) {$=$};

\node[anchor=south] at (2,0) {$\nu$};
\node[anchor=south west] at (-1,1) {$\mu$};
\node[anchor=north east] at (-1,1) {$q$};
\node[anchor=east] at (0,-2) {$\lambda$};
\node[anchor=west] at (0,-2) {$t$};
\end{tikzpicture}
\caption{Labeling associated to $C_{\lambda\mu\nu}(t,q)$.}
\label{ruleC}
\end{subfigure}
\quad
\begin{subfigure}[b]{.3\textwidth}
\centering
\begin{tikzpicture}[node distance=2cm, auto,scale=.3]
\draw[->-] (0,0) to (4,0);
\draw (0,0) to (0,-4);
\draw (0,0) to (-4,3);

\node at (0,-2) {$=$};
\node at (4,-2) {$=$};

\node[anchor=north east] at (-4,3) {$q$};
\node[anchor=south west] at (0,0) {$t$};
\node[anchor=south east] at (4,0) {$q$};
\node[anchor=east] at (6,3) {$t$};

\draw (4,0) to (7,3);
\draw (4,0) to (4,-4);

\node[anchor=south] at (2,0) {$\nu$};
\node[anchor=north east] at (0,-2) {$u_2 \downarrow$};
\node[anchor=north west] at (5,2) {$\nearrow u_1$};
\end{tikzpicture}
\caption{Two trivalent vertices glued along a non-preferred direction.}
\label{ruleP}
\end{subfigure}
\quad
\begin{subfigure}[b]{.3\textwidth}
\centering
\begin{tikzpicture}[node distance=2cm, auto,scale=.3]
\draw[->-] (0,0) to (4,0);
\draw (0,0) to (0,-4);
\node[rotate=90] at (3,0) {$=$};
\draw (0,0) to (-4,3);

\node[anchor=north east] at (-4,3) {$q$};
\node[anchor=north west] at (0,-1) {$t$};
\node[anchor=north east] at (4,-1) {$t$};
\node[anchor=east] at (6,3) {$q$};

\draw (4,0) to (7,3);
\draw (4,0) to (4,-4);

\node[anchor=south] at (2,0) {$\nu$};
\node[anchor=north east] at (0,-2) {$u_2 \downarrow$};
\node[anchor=north west] at (5,2) {$\nearrow u_1$};
\end{tikzpicture}
\caption{Two trivalent vertices glued along a preferred direction.}
\label{ruleN}
\end{subfigure}
\caption{Refined vertex conventions.
We express a leg in the preferred direction by $||$.}
\label{rule}
\end{figure}

We recall some definitions useful to work with refined topological vertex\footnote
{In subindex-packed formulas, we sometimes adopt Greek letters to label Young diagrams.}
by following the conventions used in\cite{Hayashi:2013qwa}.
An edge is labeled by a Young diagram $\nu$, and has an associated propagator $(-Q_\nu)^{|\nu|}$,
where $Q$ is the exponential of the complexified K\"ahler parameter of the corresponding 2-cycle.
Edges join at vertices, which have an associated vertex function
\be
\label{vertex-functions}
C_{\lambda\mu\nu}(t,q)
=t^{-\frac{||\mu^t||^2}2} q^{\frac{||\mu||^2+||\nu||^2}2} \tilde Z_\nu (t,q)
\sum_\eta \left(\frac{q}{t}\right)^{\frac{|\eta|+|\lambda|-|\mu|}2}
s_{\lambda^t/\eta}(t^{-\rho} q^{-\nu}) s_{\mu/\eta} (t^{-\nu^t} q^{-\rho }),
\ee
where $q=e^{-\iu \epsilon_2}$ and $t=e^{\iu \epsilon_1}$,
$s_R(q^{-\rho} t^{-\nu})$ means $s_R$ evaluated at $x_i= q^{i-\frac12}t^{-\nu(i)}$,
and $s_{\mu/\eta}(x)$ are skew-Schur functions:
if $s_\nu(x)$ is Schur function and $s_\nu(x) s_\rho(x)= \sum_\mu c^\mu_{\nu\rho} s_\mu(x)$,
then $s_{\mu/\nu}(x) := \sum_\rho c^\mu_{\nu\rho} s_\rho(x)$ \cite{macdonald-symm-func}
(in particular, $s_{\varnothing/R} =\delta_{\varnothing,R}$,
$s_{R/\varnothing} = s_R$,
and $s_{\mu/\nu}(Qq)=Q^{|\mu|-|\nu|} s_{\mu/\nu}(q)$.)
Sub-indices are ordered according to \cref{ruleC},
and we defined
\be
\label{zets}
\tilde Z_\nu(t,q)= \prod_{s \in \nu}
\left( 1-q^{l_\nu(s)} t^{a_\nu(s) +1} \right)^{-1}.
\ee
We also define
\be
\label{thefs}
f_\nu(t,q)=(-1)^{|\nu|} t^{\frac{||\nu^t||^2}2} q^{-\frac{||\nu||^2}2}, \qquad
\tilde f_\nu(t,q)=(-1)^{|\nu|} t^{\frac{||\nu^t||^2}2} q^{-\frac{||\nu||^2}2} \left(\frac{t}{q}\right)^{\frac{|\nu|}2},
\ee
so that when we glue two refined topological vertices,
we introduce framing factors $\tilde f_{\nu^t}(t, q)^n$ or $f_{\nu^t}(q, t)^n$ depending on whether
the internal line is the non-preferred direction (\cref{ruleP})
or preferred direction (\cref{ruleN}) respectively,
with $n:=\det(u_1, u_2)$.
The rule for the unrefined topological vertex is recovered by setting $t=q$.

\subsubsection{Examples}
\label{sec:examples}

\paragraph{$\SU(N)$ gauge theory}

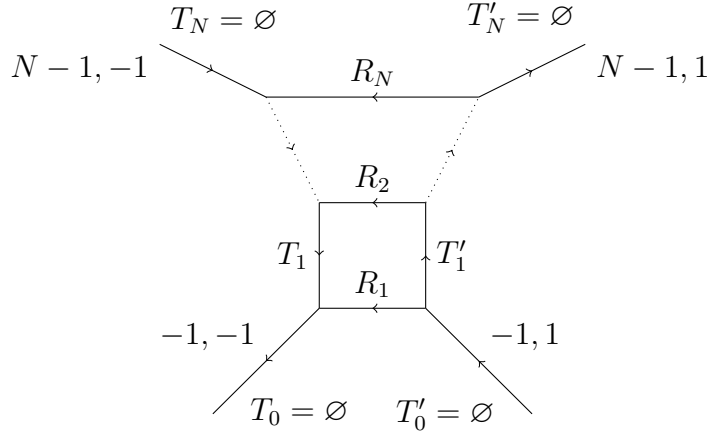
\begin{figure}[htb]
\centering

\begin{tikzpicture}[node distance=2cm, auto,scale=.7]
\draw[->-] (0,0) to (-2,-2);
\draw[->-] (2,0) to (0,0);
\draw[->-] (0,2) to (0,0);
\draw[->-] (2,0) to (2,2);
\draw[->-] (2,2) to (0,2);
\draw[->-] (4,-2) to (2,0);

\draw[dotted,->-] (-1,4) to (0,2);
\draw[dotted,->-] (2,2) to (3,4);
\draw[->-] (3,4) to (-1,4);
\draw[->-] (-3,5) to (-1,4);
\draw[->-] (3,4) to (5,5);

\node[anchor=south east] at (-1,-1) {$-1,-1$};
\node[anchor=north west] at (-1.5,-1.5) {$T_0=\varnothing$};
\node[anchor=north east] at (3.5,-1.5) {$T'_0=\varnothing$};
\node[anchor=south] at (1,0) {$R_1$};
\node[anchor=south] at (1,2) {$R_2$};
\node[anchor=south] at (1,4) {$R_N$};
\node[anchor=south west] at (3,-1) {$-1,1$};
\node[anchor=east] at (0,1) {$T_1$};
\node[anchor=west] at (2,1) {$T'_1$};
\node[anchor=north east] at (-3,5) {$N-1,-1$};
\node[anchor=south west] at (-3,5) {$T_N=\varnothing$};
\node[anchor=north west] at (5,5) {$N-1,1$};
\node[anchor=south east] at (5,5) {$T'_N=\varnothing$};

\end{tikzpicture}

\caption{Web diagram for a toric singularity engineering pure $\SU(N)$ gauge theory with $K=N-2$.}
\label{sun-oriented}
\end{figure}

As an illustrative example, consider the toric diagram for an $\SU(N)$ gauge theory, given in \cref{sun-oriented},
where we use standard notation\cite{Aganagic:2003db,Iqbal:2003zz,Iqbal:2003ix}.
Among the different possible ways to get $\SU(N)$,
we have taken our diagram to be symmetric with respect to a vertical line,
for later use in \cref{sec:examples-unoriented} when we impose $\IZ_2$ orientifold involutions.
This constrains the slope of external legs in the diagram entering the topological vertex computation
(in the gauge theory of the next section, this translates into a choice of 5d Chern-Simons level $K=N-2$ \cite{Tachikawa:2004ur}.)
We denote by $Q_{F_i}$ the exponential of the K\"ahler parameter for the edge $T_i$ (for $i=1, \ldots, N-1$).
The exponentials of the K\"ahler parameters for the horizontal edges $R_i$ are denoted by $Q_{B_i}$ (for $i=1, \ldots, N$),
and they can be expressed in terms of $Q_B := Q_{B_1} = Q_{B_2}$ as
\be
\label{theqs}
Q_{B_i} = Q_B \prod_{m=2}^{i-1} Q^{2(m-1)}_{F_m}.
\ee
The refined topological string partition function is written
\be
\label{sun-refined-vertex}
Z_\text{ref top}^{\SU(N)} &= \sum_{\substack {T_1,\ldots,T_{N-1} \\ R_1,\ldots,R_N \\ T'_1,\ldots,T'_{N-1} }}
\prod_{i=1}^N
C_{T_{i-1} T_i^t R^t_i}(t,q) \tilde f_{T_i^t}(t,q) (-Q_{F_i})^{|T_i|} \times\\
& \quad \times f_{R_i}(q, t)^{-2i+3} (-Q_{B_i})^{|R_i|} \times
C_{T'_i T'^t_{i-1} R_i}(q, t) \tilde f_{T'^t_i}(q, t) (-Q_{F_i})^{|T'_i|},
\ee
where the relevant rules and quantities are defined in \cref{sec:top-vert}.
We focus on the unrefined case (i.e.~set $t=q=e^{i\epsilon}$), where this can be explicitly evaluated to give
\be
\label{top-vertex-1}
Z_\text{top}^{\SU(N)} &= \sum_{\boldsymbol R}
\left( \prod_{i=1}^N q^{(||R_i^t||^2-||R_i||^2)(1-i) + ||R_i^t||^2} \tilde Z_{R_i}^2 Q_{B_i}^{|R_i|} \right) \times\\
&\quad\times \prod_{1 \leq k < l \leq N} \prod_{i,j=1}^\infty
\left[ 1- \left(\prod_{m=k}^{l-1} Q_{F_m}\right) q^{i+j-1-R_k(i)-R^t_l(j)} \right]^{-2}.
\ee
Using \cref{symmetric,combinatorics} we can recast the expression as sum over $N$-tuples of Young diagrams
$\boldsymbol R=(R_1,\ldots,R_N)$ such that $|\boldsymbol R|:=\sum_i |R_i|=k$,
\be
\label{top.vertex.U(N)}
Z_\text{top}^{\SU(N)} = Z_\text{top pert}^{\SU(N)} \sum_{k=0}^\infty u^k
\sum_{|\boldsymbol R|=k} CS_{N,N-2}
\prod_{i,j=1}^N \prod_{s \in R_i} \frac1{(2\iu)^2 \sin^2 \frac12 E_{ij}(s)},
\ee
where
\be
\label{Eij}
E_{ij}(s)= t_{i}- t_{j} -\epsilon_1 l_{R_i}(s) +\epsilon_2 [a_{R_j}(s)+1],
\ee
(we also denote by $E$ the unrefined expression, i.e.~the one with $\epsilon_1=-\epsilon_2 =: \epsilon$).
Here we have introduced the perturbative contribution
\be
Z_\text{top pert}^{\SU(N)}:=
\prod_{1 \leq k < l \leq N} \prod_{i,j=1}^\infty
\left[ 1- \left(\prod_{m=k}^{l-1} Q_{F_m}\right) q^{i+j-1} \right]^{-2}
\ee
and the instanton fugacity
\be
\label{fugau}
u := \frac {Q_B} {\left( \prod_{i=1}^{N-1} Q_{F_i}^{N-i} \right)^\frac2{N}}.
\ee
We have also introduced with hindsight the quantities $t_i = a_i + \iu v_i$, $i=1,\ldots, N$, satisfying $\sum_i t_i=0$,
to rephrase the K\"ahler parameters as
\be
\label{weyl-q}
Q_{F_i} = e^{ \iu (t_{i+1}- t_i)}.
\ee
$v_{i+1} - v_i$ encodes the length of vertical edge $F_i$ in the web diagram
(and eventually the gauge theory Coulomb branch parameters).
They should satisfy $|Q_{F_i}| < 1$ or $v_{i+1} > v_i$, so that it is a good expansion parameter. 
In other words, we are at a large volume point in the K\"ahler moduli space spanned by $v_i$.
We also have the quantity (eventually corresponding to the contribution from the gauge theory Chern-Simons term)
\be
\label{csdef}
CS_{N,K} := \prod_{i=1}^N \prod_{s \in R_i} e^{\iu K E_{i\varnothing}(s) }.
\ee

\paragraph{Conifold}

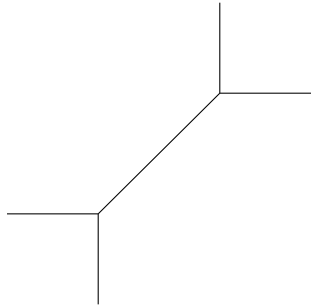
\begin{figure}[htb]
\centering
\begin{tikzpicture}[node distance=2cm, auto,scale=.4]
\draw (-2,-2) to (2,2);
\draw (2,2) to (2,5);
\draw (2,2) to (5,2);
\draw (-2,-2) to (-2,-5);
\draw (-2,-2) to (-5,-2);
\end{tikzpicture}
\caption{Web diagram for the conifold.}
\label{fig:conifold}
\end{figure}

Another illustrative example is the resolved conifold, whose web diagram
(again restricting to a case symmetric under a line reflection, for future use)
is given in \cref{fig:conifold}.
The unrefined topological vertex computation gives
\be
\label{conivertex}
Z_\text{top}^{\U(1)} =
\exp \left\{ - \sum_{m=1}^\infty \frac1m \frac{Q^m}{(q^{m/2}-q^{-m/2})^2} \right\}
= \prod_{n=1}^\infty \left( 1-Qq^{n} \right)^n,
\ee
where $Q$ is the K\"ahler parameter.

\subsection{Supersymmetric Yang-Mills on \texorpdfstring{$\epsilon$}{epsilon}-background}
\label{sec:gauge-oriented}

We now consider a 4d $\NN=2$ gauge theory.
Its exact quantum dynamics is obtained by the perturbative one-loop contribution
and the contribution from the (infinite) set of BPS instantons.
These corrections can be obtained from a 5d theory with 8 supercharges, compactified on $\IS^1$,
as a one-loop contribution from the set of 5d one-particle BPS states. 
These particles are perturbative states of the 5d theory and BPS instanton particles.

The 5d instanton partition function \cite{Nekrasov:2002qd} is given by a power series expansion
$Z_\text{inst} = \sum_{k=0}^\infty u^k Z^\text{inst}_k$, where the contribution for instanton number $k=c_2$ is
\be
\label{nek-oriented}
Z_k^\text{inst} (\{a_i\};\epsilon_1,\epsilon_2) = \tr_{\hil_k}
\left[
(-1)^F e^{-\beta H} e^{-\iu\epsilon_1 (J_1 +J_{\mathcal R})} e^{-\iu\epsilon_2 (J_2 +J_{\mathcal R})} e^{-\iu \sum a_i \Pi_i}
\right].
\ee
Here we trace over the 5d Hilbert space $\hil$ of one-particle massive BPS states.
Also, $J_1$ and $J_2$ span the Cartan subalgebra of the $\SO(4)$ little group,
$J_{\mathcal R}$ the Cartan of the $\SU(2)_{\mathcal R}$ R-symmetry,
and $\Pi$ are the Cartan generators for a gauge group $G$.
$a_i$, $i=1,\ldots,\operatorname{rank} G$ are Wilson lines of $G$ on the $\IS^1$. 

The BPS particles are W-bosons, 4d instantons (viewed as solitons in the 5d theory) and bound states thereof.
For $k \neq 0$, the partition function can be regarded as a Witten index in the SUSY quantum mechanics whose vacuum is the ADHM moduli space,
with the SUSY algebra
\be
\label{susy-alg}
\{ Q^A_M,Q^B_N \} =
P_\mu(\Gamma^\mu C)_{MN}\epsilon^{AB}+\iu\frac{4\pi^2 k}{g_\text{YM}^2}C_{MN}\epsilon^{AB}+\iu\tr(v\Pi)C_{MN}\epsilon^{AB},
\ee
where $v_i$ are the 5d Coulomb branch parameters. 
We will eventually complexify them with the already appeared Wilson lines to complete the complex Coulomb branch parameters,
bearing in mind that we will compare the Nekrasov partition function in \cref{nek-oriented}
with the topological string partition function in \cref{top.partition}.

In the following we focus on $G=\U(N)$ with 5d Chern-Simons level $K$.
The quantity in \cref{nek-oriented} can be evaluated by using localization
in equivariant K-theory \cite{Nekrasov:2002qd,  Nekrasov:2003rj, Nakajima:2005fg}
on the instanton moduli space $M(N,k)$,
more precisely on its Gieseker partial compactification and desingularization given by
framed rank $N$ torsion-free sheaves on $\CP^2=\IR^4 \cup \ell_\infty$,
where the framing is given by a choice of trivialization on the line at infinity $\ell_\infty$.
The $\epsilon$-background localizes the integral,
restricting it to a sum over
fixed points of the equivariant action.

The result of computation for pure $G=\U(N)$ gauge theory with Chern-Simons level $K$ is\cite
{Nekrasov:2002qd,Nekrasov:2003rj,Bruzzo:2002xf,Hwang:2014uwa,Tachikawa:2014dja,Tachikawa:2004ur}
\be
\label{SUN.gauge}
Z_\text{inst}^{\U(N)}=
\sum_{k=0}^\infty u^k
\sum_{|\boldsymbol R|=k} 
\prod_{i=1}^N \prod_{s \in R_i}
\frac{e^{\iu K \left(E_{i\varnothing}(s)-\frac12(\epsilon_1+\epsilon_2)\right)}}
{\prod_{j=1}^N (2\iu)^2\sin\frac{E_{ij}(s)}2 \sin\frac{E_{ij}(s)-(\epsilon_1+\epsilon_2)}2}.
\ee
For comparison with the topological string result, we take $K=N-2$.
We can restrict the result to $\SU(N)$ by constraining the sum of the Coulomb branch moduli to be zero.
Note that the $\U(N)$ and $\SU(N)$ instanton partition functions are in general different,\footnote
{The difference between the $\U(N)$ Nekrasov partition functions and the $\SU(N)$ Nekrasov partition functions
has been discussed in \cite{Bergman:2013ala, Bao:2013pwa, Hayashi:2013qwa, Bergman:2013aca, Hwang:2014uwa}.
It turns out that the web diagram nicely encodes factors that account for the difference.}
but they agree in our case of zero flavors with Chern-Simons level $K=N-2$.
With this proviso, we can see that \cref{top.vertex.U(N)} from \cref{sec:examples}
can be written as $Z_\text{top}^{\SU(N)} / Z_\text{top pert}^{\SU(N)}= Z_\text{inst}^{\U(N)}$
evaluated for $K=N-2$, $\sum t_i=0$,
and in the unrefined limit $\epsilon_1=-\epsilon_2 =: \epsilon$, by identifying the complexified K\"ahler parameter with the complexified Coulomb branch moduli. 
Hence we have an exact match up to the perturbative part.

\medskip

Another interesting example  is $\U(1)$ gauge theory.
Although it does not support semi-classical gauge instantons, one can consider 
BPS states corresponding to small instantons.
A mathematically more rigorous way to define them is to consider
$\U(1)$ instantons on non-commutative $\IR^4$, or equivalently rank 1 torsion-free sheaves on $\CP^2$
with fixed framing on the line at infinity\cite{Nekrasov:1998ss}.
The gauge theory result is\cite{Nakajima:2003pg,Szendroi:2012ke}
\be
\label{k-theor}
\sum_R \frac{Q^{|R|}}
{\prod_{s \in R} \left( 1- q_1^{-l_R(s)}q_2^{1+a_R(s)}\right) \left( 1- q_1^{1+l_R(s)}q_2^{-a_R(s)}\right)}=
\exp \left\{ \sum_{r=1}^\infty \frac1r \frac{Q^r}{(1-q_1^r)(1-q_2^r)} \right\},
\ee
where $q_1=e^{\iu\epsilon_1}$, $q_2=e^{\iu\epsilon_2}$.
The exponent agrees precisely with the topological vertex result \cref{conivertex} in the unrefined case $q_1q_2=1$,
by setting $q_1=q$.

\section{Orientifolds and the real topological vertex}
\label{sec:unoriented} 

In this section we review properties of the unoriented theories we are going to focus on. We first introduce their description in string theory and M-theory, and subsequently review the computation of their partition function using the real topological string theory in the real topological vertex formalism.

\subsection{Generalities}
\label{sec:gen-unoriented}

There are many ways to obtain an unoriented theory from a parent oriented string theory configuration,
which in our present setup result in different gauge theory configurations.
In this paper we will focus on a particular choice,
which has the cleanest connection with the parent oriented theory, in a sense that we now explain.

Consider the type IIA version of our systems, namely type IIA on a non-compact toric CY threefold $\IX$ singularity.
We introduce an orientifold quotient, acting as an antiholomorphic involution $\sigma$ on $\IX$
and as a sign flip in an $\IR^2$ (parametrized by $x^2,x^3$) of 4d Minkowski space.
For concreteness, we consider $\sigma$ to have a fixed locus $L$, which on general grounds is a lagrangian 3-cycle of $\IX$
(one can build orientifolds with similar M-theory lift even if $\sigma$ is freely acting).
In other words, we have an O4-plane wrapped on $L$ and spanning $x^0,x^1$;
we choose the O4-plane to carry negative RR charge (see later for other choices).
We complete the configuration by introducing one single D4-brane (as counted in the covering space)
wrapped on $L$ and spanning $x^0,x^1$, namely stuck on the O4$^-$-plane.\footnote
{Additional pairs of mirror D4-branes may be added;
in the M-theory lift, they correspond to the inclusion of explicit M5-branes, so that open M2-brane states enter the computations.}

In general, if $H_1(L;\IZ)$ is non-trivial, it is possible to turn on $\IZ_2$-valued Wilson lines for the D4-branes worldvolume
$\mathrm O(1)\equiv \IZ_2$ gauge group.
This results in a different sign weight for the corresponding disk amplitudes, as discussed in the explicit example later on.

This setup is the physical realization of the real topological string introduced in \cite{Walcher:2007qp}
(see also \cite{Krefl:2008sj,Krefl:2009md,Krefl:2009mw}).
In the topological setup, the addition of the D4-brane corresponds to a topological tadpole cancellation condition;
in the physical setup, it corresponds to local cancellation of the RR charge, and leads to a remarkably simple M-theory lift,
which allows for direct connection with the 5d picture of the oriented case, as follows.

This IIA configuration lifts to M-theory as a compactification on $\IX \times \IS^1$,
with a $\IZ_2$ quotient\footnote
{The M-theory 3-form $C_3$ is intrinsically odd under this $\IZ_2$, so we refer to it as `orientifold action' in M-theory as well.}
acting as $\sigma$ on $\IX$, as $(x^2,x^3)\to (-x^2,-x^3)$ on 4d Minkowski space-time, and as a half-shift along the $\IS^1$.
Because the $\IZ_2$ is freely acting on the $\IS^1$,
the configuration can be regarded as an $\IS^1$ compactification of the 5d theory corresponding to M-theory on $\IX$
(with the $\IS^1$ boundary conditions for the different fields given by their eigenvalue under the orientifold action).
Since the 5d picture is essentially as in the oriented case,
these configurations have a direct relation with the oriented Gopakumar-Vafa description of the topological string.
Specifically, the real topological string amplitude is given by a one-loop diagram of 5d BPS states running on $\IS^1$,
with integer (resp.~half integer) KK momentum for states even (resp.~odd) under the orientifold action \cite{unoriented-gv}.

For completeness, we quote the M-theory lifts corresponding to other choices of O4-plane and D4-brane configurations \cite{Hori:1998iv,Gimon:1998be}:
\begin{itemize}
\item An orientifold introducing an O4$^-$-plane with no stuck D4-brane lifts to M-theory on $\IX \times \IS^1$
with a $\IZ_2$ acting as $\sigma$ on $\IX$, flipping $x^2,x^3$ in 4d space, and leaving the $\IS^1$ invariant.
This M-theory configuration has orbifold fixed points
and therefore is not directly related to the 5d picture of the oriented theory.

\item An orientifold introducing an O4$^+$-plane lifts to M-theory on $\IX \times \IS^1$
with a $\IZ_2$ acting as $\sigma$ on $\IX$, flipping $x^2,x^3$ in 4d space, and leaving the $\IS^1$ invariant,
with 2 M5-branes stuck at the orbifold locus.
Again this M-theory configuration has orbifold fixed points.

\item Finally, there is an exotic orientifold, denoted $\widetilde{\text{O}4^+}$-plane,
which lifts to M-theory as our above freely acting orbifold (acting with a half-shift on $\IS^1$),
with one extra stuck M5-brane.
This M-theory configuration contains a sector of closed membranes exactly as in the O4$^-$+D4 case,
and in addition an open membrane sector which has no direct relation to the 5d oriented theory
(but is described by Ooguri-Vafa invariants \cite{Ooguri:1999bv}).

\end{itemize}

Hence, as anticipated, we focus on the O4$^-$+D4,
whose M-theory lift is the simplest and closest to the parent 5d oriented theory.

\medskip

\begin{figure}[htb]
\centering

\begin{tikzpicture}[node distance=2cm, auto,scale=.7]
\draw (2,1) -- (2,-1) -- (-2,-1) -- (-2,1) -- (2,1);
\draw (2,1) -- (3,3) -- (-3,3) -- (-2,1) -- (2,1);
\draw (2,-1) -- (3,-3) -- (-3,-3) -- (-2,-1) -- (2,-1);
\draw (3,3) to (6,4);
\draw (-3,3) to (-6,4);
\draw (-3,-3) to (-6,-4);
\draw (3,-3) to (6,-4);

\draw[red,dashed] (0,-4.5) to (0,4.5);
\draw[blue,dashed] (-6.5,0) to (6.5,0);
\end{tikzpicture}

\caption{$\SU(4)$ with two involutions.}
\label{su4}
\end{figure}
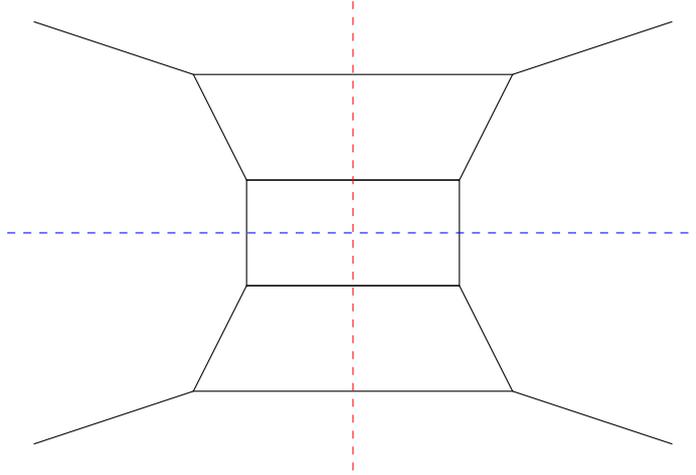

To finally determine the orientifold actions, we must specify the antiholomorphic involution $\sigma$ acting on $\IX$.
In general, a toric CY three-fold associated to an $\SU(N)$ gauge theory admits two such $\IZ_2$ actions,\footnote
{Certain cases, like the conifold, may admit additional symmetries.}
illustrated in \cref{su4} for $\SU(4)$.
They mainly differ in the effect of the orientifold action on the Coulomb branch moduli of the 5d gauge theory. Namely, the blue quotient in \cref{su4} reduces the number of independent moduli,
whereas the red one preserves this number.
Equivalently, the two quotients either reduce or preserve the rank of the gauge group at the orientifold fixed locus.
Since the Coulomb branch parameters play an important role in the parent gauge theory localization computation,
we will focus on rank-preserving quotients to keep the discussion close to the parent theories.
We leave the discussion of rank-reducing involutions for future projects.

\subsection{The real topological string}
\label{sec:top-unoriented}

The real topological string is a natural generalization of the topological string in \cref{sec:top-oriented}.
It provides a topological version of the IIA orientifolds in the previous section.
Namely, the real topological string computes holomorphic maps from surfaces with boundaries and crosscaps into
a target $\IX$ modded out by the orientifold involution $\sigma$.
Realizing the unoriented world-sheet surface as a quotient of a Riemann surface by an antiholomorphic involution,\footnote
{The case $\Sigma$ itself is a Riemann surface requires to start from a disconnected $\Sigma_g$,
and is better treated separately.}
$\Sigma=\Sigma_g/\Omega$,
we must consider equivariant maps $f$ as in \cref{comm-diag}.
\begin{figure}[htb]
\centering
\begin{tikzpicture}[node distance=2cm, auto,scale=.7]
  \node (P) at (-1,1) {$\Sigma_g$};
  \node (B) at (1,1) {$\Sigma_g$};
  \node (A) at (-1,-1) {$\IX$};
  \node (C) at (1,-1) {$\IX$}; 
  \node (D) at (0,0) {$\circlearrowleft$};
  \draw[->] (P) to node {$\Omega$} (B);
  \draw[->] (P) to node [swap] {$f$} (A);
  \draw[->] (A) to node [swap] {$\sigma$} (C);
  \draw[->] (B) to node {$f$} (C);
\end{tikzpicture}
\caption{Commutative diagram for equivariant map.}
\label{comm-diag}
\end{figure}

The model includes crosscaps, and boundaries
(with a single-valued Chan-Paton index to achieve the topological tadpole cancellation)
ending on the lagrangian $L$.
Hence, we must consider the relative homology class $f_* ([\Sigma_g]) \in H_2(\IX,L;\IZ)$.

The total topological amplitude at fixed Euler characteristic $\chi$ may be written as
\be
\mathcal G^{(\chi)}=\frac12
\left[ \FF ^{(g_\chi)}+ \sum \FF ^{(g,h)} + \sum \mathcal R^{(g,h)} + \sum \mathcal K^{(g,h)} \right],
\ee
where the different terms account for closed oriented surfaces, oriented surfaces with boundaries,
surfaces with one crosscap, and surfaces with two crosscaps.
Different consistency conditions,
needed to cancel otherwise ill-defined contributions from the enumerative geometry viewpoint,\footnote
{For example, for configurations in which $H_1(L;\IZ)=\IZ_2$ and $H_2(\IX,L;\IZ)=\IZ$,
one needs to cancel homologically trivial disks against crosscaps.}
guarantee integrality of the BPS expansion for
\be
\FF_\text{real} = \sum_\chi \iu^\chi \epsilon^\chi \left( \mathcal G^{(\chi)} - \frac12 \FF ^{(g_\chi)}\right).
\ee
This can be taken as the definition of the real topological string.

This integrality of BPS invariants,
as well as a physical explanation of the tadpole cancellation and other consistency conditions of the real topological string,
may be derived from the M-theory viewpoint\cite{unoriented-gv}.
The real topological string amplitude is obtained as a sum over 5d BPS M2-brane states of the oriented theory,
running in the compactification $\IS^1$ with boundary conditions determined by the eigenvalue under the orientifold operator.
For a short review, see \cref{sec:m-theory}.
Denoting by $\hgv$ this weighted BPS multiplicity of M2-branes wrapped on a genus $\hg$ surface
(as counted in the quotient)
in the homology class $\beta$, the equivalent to \cref{gv-closed-d} is
\be
\label{final-trace}
\FF_\text{real} =
\sum_{\substack {\beta,\hg \\ \text{odd } m \geq 1} } \hgv \, \frac 1m
\left[ \, 2\sinh \left( \frac {m\epsilon}2 \right) \,\right] ^{\hg-1}
e^{\iu m\beta\cdot t}.
\ee

To compute real topological string partition function on local Calabi-Yau,
we will use the real topological vertex \cite{Krefl:2009mw},
which is a generalization of the standard topological vertex to take into account involutions of the toric diagram.
The formalism is still only available in the unrefined case, on which we focus herefrom.

\begin{figure}[htb]
\centering
\begin{tikzpicture}[node distance=2cm, auto,scale=.4]
\draw (-2,-2) to (2,2);
\draw (2,2) to (2,5);
\draw (2,2) to (5,2);
\draw (-2,-2) to (-2,-5);
\draw (-2,-2) to (-5,-2);

\draw[red,dashed] (-3,3) to (3,-3);
\end{tikzpicture}
\qquad
\begin{tikzpicture}[node distance=2cm, auto,scale=.4]
\draw (-2,2) -- (2,2) -- (2,-2) -- (-2,-2) -- (-2,2);
\draw (2,2) to (4,4);
\draw (-2,2) to (-4,4);
\draw (2,-2) to (4,-4);
\draw (-2,-2) to (-4,-4);

\draw[red,dashed] (0,3) to (0,-3);
\end{tikzpicture}
\caption{Web diagram for our orientifolds of the conifold and $\SU(2)$ theories.}
\label{fig:ori-coni-sutwo}
\end{figure}
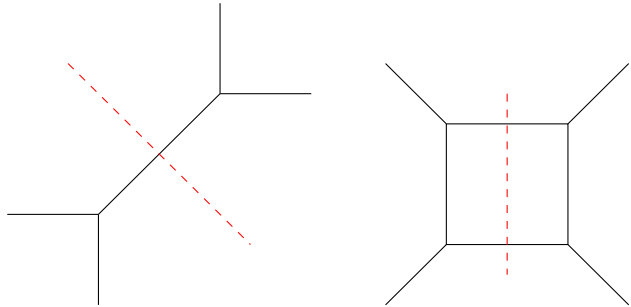

We apply the formalism to involutions of the kind shown in \cref{fig:ori-coni-sutwo},
as described in more detail in \cref{sec:real-vertex}.
For these involutions there are no legs fixed point-wise in the diagram,
and this simplifies the computation of the topological vertex.
Due to the symmetry of the diagram in the parent theory,
one can use symmetry properties of the vertex functions to cast each summand in the sum over Young diagrams as a square\cite{Krefl:2009mw}.
Then the real topological vertex amplitude is given by the sum of the square roots of the summands.
To define these in a consistent way, we follow the choice of sign in \cite{Krefl:2009mw}, 
see \cref{sign.real}. In all our examples this sign is trivial, since $|R| \pm c(R)$ is even for every $R$, and in the cases we consider also $n+1=-2i+4$ is even as well, as can be seen from \cref{sun-refined-vertex}.

Explicit examples will be described in \cref{sec:examples-unoriented}.

\section{Explicit examples}
\label{sec:examples-unoriented}

In this section, we explicitly compute the real topological string partition functions of the resolved conifold and also the $\SU(N)$ geometry using the real topological vertex formalism\cite{Krefl:2009mw}. We first review the calculation of the real topological string partition function of the resolved conifold\cite{Krefl:2009mw}, correcting some typos. Then, we move on to the computation of the real topological string partition function for the $\SU(N)$ geometry, which shows an intriguing new feature.

\subsection{The real conifold} 
\label{sec:coni-unoriented}

Let us apply the above recipe to the orientifold of the resolved conifold.
This is particularly simple because there are no Coulomb branch moduli,
and the only parameters are the instanton fugacity and those defining the $\epsilon$-background.

The topological string side can be computed using the real topological vertex formalism.
The result\footnote{This expression corrects some typos in\cite{Krefl:2009mw}.} reads
\be
\label{real-conifold-vertex}
Z_\text{real top} ^{\U(1)}= \exp \left\{ - \frac12 \sum_{m=1}^\infty \frac1m \frac{Q^m}{(q^{m/2}-q^{-m/2})^2}
\pm \sum_{m \text{ odd}} \frac1m \frac{Q^{m/2}}{q^{m/2}-q^{-m/2}}\right\}.
\ee
Our choice of orientifold plane charge corresponds to the negative sign.\footnote{The choice of positive sign can be recovered by turning on a non-trivial $\IZ_2$-valued Wilson line on the D4-brane stuck at the O4-plane in the type IIA picture, since the fixed locus $L=\IR^2\times \IS^1$ has one non-trivial circle.}

The first term in the exponent corresponds to the closed topological string contribution, while the second 
reproduces the open and unoriented topological string contributions.

\subsection{Orientifold of pure \texorpdfstring{$\SU(N)$}{SU(N)} geometry, and its flops}
\label{sec:sun-unoriented}

In this section we study the unoriented version of the $\SU(N)$ systems of \cref{sec:examples}, compute their real topological vertex amplitudes following the rules in \cite{Krefl:2009mw}, and describe their behavior under flops of the geometry. 

\subsubsection{Real topological vertex computation}
\label{sec:real-top-vert-sun}

\begin{figure}[htb]
\centering

\begin{tikzpicture}[node distance=2cm, auto,scale=.7]
\draw[->-] (0,0) to (-2,-2);
\draw[->-] (2,0) to (0,0);
\draw[->-] (0,2) to (0,0);
\draw[->-] (2,0) to (2,2);
\draw[->-] (2,2) to (0,2);
\draw[->-] (4,-2) to (2,0);

\draw[dotted,->-] (-1,4) to (0,2);
\draw[dotted,->-] (2,2) to (3,4);
\draw[->-] (3,4) to (-1,4);
\draw[->-] (-3,5) to (-1,4);
\draw[->-] (3,4) to (5,5);

\draw[dashed,red] (1,5) to (1,-2);

\node[anchor=south east] at (-1,-1) {$-1,-1$};
\node[anchor=north west] at (-1.5,-1.5) {$T_0=\varnothing$};
\node[anchor=north east] at (3.5,-1.5) {$T'_0=\varnothing$};
\node[anchor=south] at (1,0) {$R_1$};
\node[anchor=south] at (1,2) {$R_2$};
\node[anchor=south] at (1,4) {$R_N$};
\node[anchor=south west] at (3,-1) {$-1,1$};
\node[anchor=east] at (0,1) {$T_1$};
\node[anchor=west] at (2,1) {$T'_1$};
\node[anchor=north east] at (-3,5) {$N-1,-1$};
\node[anchor=south west] at (-3,5) {$T_N=\varnothing$};
\node[anchor=north west] at (5,5) {$N-1,1$};
\node[anchor=south east] at (5,5) {$T'_N=\varnothing$};

\end{tikzpicture}

\caption{$\SU(N)$ with $K=N-2$ and involution.}
\label{sun}
\end{figure}
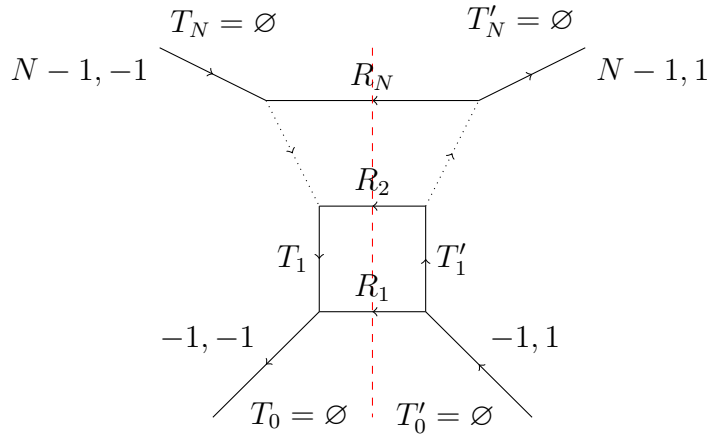

The web diagram is given in \cref{sun}, which describes a $\IZ_2$ involution of \cref{sun-oriented}
(which was chosen symmetric in hindsight).

We recall some expressions already introduced in \cref{sec:examples} for the oriented case.
We define the perturbative contribution as
\be
Z_\text{real top pert}^{\SU(N)}:=
\prod_{1\leq k < l \leq N} \prod_{i,j=1}^\infty \left( 1- \left(\prod_{m=k}^{l-1}Q_{F_m}\right) q^{i+j-1} \right)^{-1}.
\ee
We also recall the Chern-Simons level \cref{csdef}
\be
CS_{N,K} := \prod_{i=1}^N \prod_{s \in R_i} e^{\iu K E_{i\varnothing}(s) },
\ee
where $E_{ij}$ is defined in \cref{Eij} and $\varnothing$ denotes the empty diagram.
Finally, we introduce the rescaled instanton fugacity
\be
\tilde u := \frac {Q_B^\frac12} {\left( \prod_{i=1}^{N-1} Q_{F_i}^{N-i} \right)^\frac1{N}},
\ee
which is the square root of the instanton fugacity \cref{fugau} in the oriented computation.

Expressing the (complexified) K\"ahler parameters
in terms of the (complexified) edge positions $t_i$, with $\sum_i t_i =0$,
and edges ordered such that $\operatorname{Im} t_{i+1} > \operatorname{Im} t_{i}$ in a certain large volume region in the  K\"ahler moduli space, we take as in \cref{weyl-q}
\be
\label{kahlercoulomb}
Q_{F_i} = e^{\iu(t_{i+1}- t_i)}.
\ee
Notice that in this case the fugacity can be written as
$\tilde u = Q_B^\frac12 e^{\iu t_1}$.

The computation is as follows: we start from \cref{sun-refined-vertex}, go to the unrefined limit,
and apply real topological vertex rules\cite{Krefl:2009mw}, cf.~\cref{sec:real-vertex}.
Since our involution does not fix any leg point-wise,
we only need to reconstruct the square within summands
using permutation properties of the topological vertex $C_{RR'R''}(q)$ (see \cref{permutation}),
and the fact that $T_i=T'_i$ due to the involution.
We get
\be
\left.Z_\text{top}^{\SU(N)}\right|_{T_i=T'_i}=
\sum_{\substack {T_1,\ldots,T_{N-1} \\ R_1,\ldots,R_N }} \prod_{i=1}^N
C^2_{T_i T_{i-1}^t R_i}(q)
Q_{F_i}^{2|T_i|}Q_{B_i}^{|R_i|}
(-1)^{|R_i|(4-2i)} \\
q^{||T_i||^2-||T_i^t||^2+(2-i)\left(||R_i^t||^2-||R_i||^2\right)}.
\ee
We then take the square root, and notice that the sign \cref{sign.real} for the propagator is always $+1$,
\be
Z_\text{real top}^{\SU(N)} &=
\sum_{\substack {T_1,\ldots,T_{N-1} \\ R_1,\ldots,R_N}}
\prod_{i=1}^N
C_{T_i T^t_{i-1} R_i}(q)
Q_{F_i}^{|T_i|}  Q_{B_i}^{|R_i|/2} \\
& q^{\frac14 (||T_i||^2-||T^t_i||^2) + \frac14 (||T_{i-1}||^2-||T^t_{i-1}||^2) + \frac{2-i}2 (||R_i^t||^2-||R_i||^2)}.
\ee
By using combinatorial identities for Young diagrams and skew-Schur functions,
described in \cref{sec:help} (in particular \cref{schur-id,symmetric}), we arrive at the final result
\be
\label{vertex-su-n}
Z_\text{real top}^{\SU(N)} = Z_\text{real top pert}^{\SU(N)} \sum_{k=0}^\infty \tilde u^k
\sum_{|\boldsymbol R|=k} (-1)^{\sum_i (i-1) |R_i|}
CS_{N,\frac{N-2}2}
\prod_{i,j=1}^N \prod_{s \in R_i} \frac1{2\iu\sin \frac12 E_{ij}(s)}.
\ee

\subsubsection{Behavior under flops}
\label{sec:flops}

It is worth to briefly step back and emphasize an important point.
In the above computation there is an explicit choice of ordering of edges in the web diagram,
which defines a particular large volume limit.
Moving in the K\"ahler moduli space across a wall of a flop transition\footnote
{The flop transition considered here shrinks and introduces a family of rational curves. This is different from the usual flop which shrinks and introduces an isolated rational curve. The behavior of the (non-real) topological string partition function under the usual flop has been studied in\cite{Iqbal:2004ne, Konishi:2006ev, Taki:2008hb}.}
can reorder the edges, so we need to redefine the expansion parameters \cref{weyl-q}.
Consider the simplest setup in which the ordering of all edges is reversed,
such that $v_{i}> v_{i+1}$ for all $i$,
and we take
\be
Q_{F_i} = e^{\iu (t_{N-i}-t_{N-i+1})},
\ee
with $|Q_{F_i}| < 1$.
In this case, we are at a different large volume point in the enlarged K\"ahler moduli space
compared to the case when we defined $t_i$ by \cref{kahlercoulomb}. 
From the viewpoint of the five-dimensional pure $\SU(N)$ gauge theory, it corresponds to moving to a different Weyl chamber in the Coulomb branch moduli space by a Weyl transformation $t_i \rightarrow t_{N-i+1}$ for all $i$. 
If we write the result by using $E_{ij}(s)$ and $CS_{N,K}$ defined in \cref{Eij,csdef},
this gives the same exact result \emph{except} for sign pattern $(-1)^{\sum_i (N-i)|R_i|}$.
The computation of this result is similar to \cref{vertex-su-n} except that
we redefined dummy variables $R_i^\text{new}:=(R_{N-i+1}^{\text{old}})^t$, compared to the geometry before the flop transition.
When we regard \cref{vertex-su-n} as a function of $t_i$, we have computed
\be
Z_\text{real top}(t_{N-i+1}, K) =: \tilde Z_\text{real top}(t_i, K),
\ee
which is not equal to $Z_\text{real top}(t_i, K)$ as a function of $t_i$ in the unoriented case.\footnote
{In the case $N=3$ they are equal to each other accidentally.}
This is different from the oriented case where we have $Z_\text{top} (t_i, K) = Z_\text{top} (t_{N-i+1}, K)$,
which should be true since Weyl transformations are part of the gauge transformations.
Therefore, this is a feature special to the real topological string partition function of pure $\SU(N)$ geometry. 
From the viewpoint of five-dimensional pure $\SU(N)$ gauge theory,
the pure $\SU(N)$ gauge theory is invariant under the Weyl transformation of $\SU(N)$
and this is reflected into the invariance of the partition function under the Weyl transformation in the oriented case. 
In the unoriented case, however, the non-invariance of the partition function under the transformation
implies that the presence of the orientifold or the corresponding defect in field theory
breaks the symmetry that existed in the oriented case.

It is similarly easy to consider intermediate cases of partial reorderings.
The simplest is to take $N=3$, and move from $v_1< v_2 < v_3$ to $v_2 < v_3 < v_1$.
In this case, the new expansion parameters are $Q_{F_1}=e^{\iu (t_3 - t_2)}$ and $Q_{F_2}=e^{\iu( t_1 - t_3)}$.
The result is basically the same, but with sign $(-1)^{2|R_1|+|R_3|}$.
Here we redefined dummy variables as
$R^\text{new}_1 :=( R^\text{old}_3)^t$, $R^\text{new}_2 := (R^\text{old}_1)^t$, $R^\text{new}_3 := (R^\text{old}_2)^t$.

In other words, starting with the result in a given chamber,
moving across a wall of a flop transition exchanging two edges with diagrams $R$, $S$
produces a change in the amplitude (expressed in the new K\"ahler parameters) given by a sign $(-1)^{|R|+|S|}$.

This is the explicit manifestation of the fact that the topological string amplitude regarded as a function of $t_i$
in this unoriented theory is not universal throughout the moduli space, but it has a non-trivial behavior.\footnote
{However, when we regard the topological string amplitude as a function of a good expansion parameter
which is always $Q_{F_i}$ with $|Q_{F_i}| < 1$, then they are essentially the same function.
Namely, the real GV invariants are the same at the two different points in the enlarged K\"ahler moduli space.}

\subsubsection{Behavior under other transformations}

Let us consider another transformation which is a refection with respect to a horizontal axis
for the pure $\SU(N)$ geometry of \cref{sun}.
The operation in the original pure $\SU(N)$ geometry corresponds to charge conjugation,
which is given by a transformation of the Coulomb branch moduli $t_i \rightarrow -t_{N-i+1}$ for $i=1, \cdots, N$ and a flip of the sign of CS level.\footnote
{When we regard the pure $\SU(N)$ geometry as a 5-brane web diagram, the CS level can be read off from the asymptotic behavior of the external legs\cite{Bergman:1999na, Kitao:1998mf, Bergman:2013aca}. In particular, when we turn the 5-brane web upside-down, the sign of the CS level of the gauge theory also flips.}
The transformation can be effectively implemented by defining the K\"ahler parameters as
\be
\label{kahlercoulomb2}
Q_{F_i} = e^{\iu (t_i-t_{i+1})},
\ee
with $|Q_{F_i}| < 1$ and the same labeling for $R_i$ for all $i$ as in \cref{sun}.
Compared to \cref{kahlercoulomb}, we flip the sign of $t_i$ for all $i$.
Hence, we are now assuming $v_{i} > v_{i+1}$ for all $i$ and hence we effectively consider the pure $\SU(N)$ geometry upside-down.
If we write the result by using $E_{ij}(s)$ and $CS_{N,K}$ defined in \cref{Eij,csdef},
this gives the same result except for different CS level $-\frac{N-2}2$ and sign pattern $(-1)^{\sum_i (N-i)|R_i|}$.
In this case, we did not redefine the dummy variables $R_i$.
The change in the sign of CS level is consistent with the fact that the
definition \cref{kahlercoulomb2} is related to charge conjugation of the original pure $\SU(N)$ gauge theory.
When we regard \cref{vertex-su-n} as a function of $t_i$, we have computed
\be
Z_\text{real top}(-t_i, -K) =: \tilde{Z}_\text{real top}(t_i, K),
\ee
which is again not equal to $Z_\text{real top}(t_i, K)$ as a function of $t_i$ in the unoriented case.
This is different from the oriented case where we have $Z_{\text{top}}(t_i, K) = Z_{\text{top}}(-t_i, -K)$.\footnote
{We have checked this for $N=2,3,4$ up to 5-instanton order.}
This is another example of a transformation where the presence of the orientifold defect breaks the invariance of the partition function under a transformation that existed in the original theory without the defect.

\section{Discussion: towards gauge theory interpretation}
\label{sec:towards-gauge}

As explained in the introduction, the real topological string on the local CY threefold should be related to the partition function of the corresponding gauge theory in the presence of a surface defect. Given the M-theory lift of the orientifold in terms of a freely-acting shift on the $\IS^1$, this should correspond to a partition function of the theory on $\IS^1$, with modified boundary conditions, or equivalently with an extra twist in the Witten index computation,
\be
\label{nek-unoriented}
Z_k^\text{real inst} (\{a_i\};\epsilon_1,\epsilon_2) = \tr_{\hil_k}
\left[(-1)^F e^{-\beta H} e^{-\iu\epsilon_1(J_1 +J_{\mathcal R})} e^{-\iu\epsilon_2 (J_2 +J_{\mathcal R})} e^{-\iu \sum a_i \Pi_i} \mathcal O_\Omega
\right],
\ee
where $\mathcal O_\Omega$ is an operator implementing the orientifold action in the corresponding Hilbert space sector.

In this section we exploit the intuitions from the topological vertex computations to describe aspects of this twist in explicit examples.

\subsection{Invariant states and the conifold example}
\label{sec:invariant}

We start the discussion with the conifold. This is particularly simple, because there is only one BPS (half-)hypermultiplet, whose internal structure is invariant under the orientifold, namely it is an M2-brane wrapped on a 2-cycle mapped to itself under the orientifold. Then the orientifold action is just action on the Lorentz quantum numbers. From the viewpoint of gauge theory, there is a 5d $\U(1)$ gauge theory, whose BPS states are instantons. In this simple system it is possible to motivate the structure of the twisted Nekrasov partition function \cref{nek-unoriented}, i.e.~of the operator $\mathcal O_\Omega$. As discussed in \cref{sec:examples}, the ADHM moduli space is just $(\IC^2)^k/S_k$ where $k$ is the instanton number and $S_k$ is the symmetric group of order $k$; these moduli are intuitively the positions of the $k$ instantons in $\IR^4$. Therefore the action of $\mathcal O_\Omega$ on this moduli space is simply the geometric action imposed by the orientifold. 
Since the orientifold action flips the space-time coordinates $(x^2,x^3)\to (-x^2,-x^3)$, we are motivated to take $O_\Omega$ as given by a shift
\be
\epsilon_2 \rightarrow \epsilon_2+\pi
\ee
in the original parent gauge theory expression \cref{nek-oriented}. Furthermore, we assume that the operator induces the redefinition by a factor of 2 of certain quantities between the parent theory and the twisted theory.
In practice, it requires that the twisted theory result should be expressed in terms of the redefined weight
\be
Q \to Q^\frac12.
\ee

We then consider the twisted Nekrasov partition function of the $\U(1)$ instanton. First, we consider the refined amplitude for the original theory \cref{k-theor}, and perform the shift $\epsilon_2 \to \epsilon_2 +\pi$. Taking the unrefined limit, we obtain
\be
\label{unrefined-conifold-gauge}
\exp \left\{ -\frac12 \sum_{m=1}^\infty \frac1m \frac{Q^{2m}}{(2\iu\sin\epsilon m)^2}
- \sum_{k \text{ odd}} \frac1k \frac{Q^k}{2\iu\sin \epsilon k} \right\}.
\ee
We now redefine $\epsilon = \tilde\epsilon/2$ and $Q=\tilde Q^\frac12$, and get
\be
Z_\text{real inst} ^{\U(1)}= \exp \left\{ -\frac12 \sum_{m=1}^\infty \frac1m \frac{\tilde Q^m}{(2\iu\sin\frac12 \tilde\epsilon m)^2}
- \sum_{k \text{ odd}} \frac1k \frac{\tilde Q^{k/2}}{2\iu\sin \frac12 \tilde \epsilon k} \right\},
\ee
which agrees with \cref{real-conifold-vertex} for the negative overall sign for the unoriented contribution. One may choose a redefinition $Q = -\tilde Q^\frac12$, which agrees with \cref{real-conifold-vertex} for the positive overall sign for the unoriented contribution. This choice reflects the choice of $\IZ_2$ Wilson line, 
although its gauge theory interpretation is unclear.

Note that the conifold geometry is special in that the only degree of freedom is one BPS (half-)hypermultiplet, from an M2-brane on a $\CP^1$ invariant under the orientifold action, which thus motivates a very simple proposal for $\mathcal O_\Omega$. This can in general change in more involved geometries, where there are higher spin states, and/or states not invariant under the orientifold. In these cases, the orientifold action should contain additional information beyond its action on 4d space-time quantum numbers.

\subsection{A twisted Nekrasov partition function for pure \texorpdfstring{$\SU(N)$}{SU(N)}}
\label{gauge-sun-comp}

We now consider the case of the $\SU(N)$ geometry, whose real topological string amplitude was described in \cref{sec:real-top-vert-sun}, and discuss its interpretation in terms of a twisted Nekrasov partition function.

We start with the following observation. Consider the parent theory expression \cref{SUN.gauge} for CS level $K = N-2$ as starting point. Since part of the orientifold action includes a space-time rotation which motivates the $\epsilon_2$-shift, let us carry it out just like in the conifold case and check the result.

Let us thus perform the $\epsilon_2$-shift and take the unrefined limit
($\epsilon_2 \to -\epsilon + \pi$, $\epsilon_1 \to \epsilon$),
and rescale $\epsilon\rightarrow \frac{\epsilon}2$, and $t_i \rightarrow \frac{t_i}2$
with $u \rightarrow \tilde u (=u^\frac12)$.
We obtain the result
\be
\label{SUN.gauge.real.i}
Z_\text{real inst}^{\U(N)}
=\sum_{k=0}^\infty \tilde u^{k}\sum_{|{\boldsymbol R}|=k} 
(-1)^{k+A+\Gamma} \prod_{i=1}^N \prod_{s \in R_i}
\frac{e^{\iu \frac{N-2}2 E_{i\varnothing}(s)}}{\prod_{j=1}^N(2\iu)\sin\frac12 E_{ij}(s)},
\ee
where $A=\frac12 N\left( \sum_i ||R_i^t||^2+ |R_i|\right)$, and $\Gamma = \sum_{ij} \sum_{s \in R_i} a_{R_j}(s)$.
One can prove that $A+\Gamma = \sum_{ij} \sum_{(m,n)\in R_i} R_j^t (n) \equiv k \pmod{2}$,
so the final result is
\be
\label{SUN.gauge.real}
Z_\text{real inst}^{\U(N)}
=\sum_{k=0}^\infty \tilde u^{k}\sum_{|{\boldsymbol R}|=k}  
\prod_{i=1}^N \prod_{s \in R_i}
\frac{e^{\iu \frac{N-2}2 E_{i\varnothing}(s)}}{\prod_{j=1}^N(2\iu)\sin\frac12 E_{ij}(s)}.
\ee
This expression evaluated for $\sum t_i=0$ is remarkably close to the real topological string computation \cref{vertex-su-n},
up to $i$-dependent sign factors. Specifically \cref{vertex-su-n} can be recast as
\be
\label{signed-fmla}
Z_\text{real top}^{\SU(N)} / Z_\text{real top pert}^{\SU(N)}
=\sum_{k=0}^\infty \tilde u^{k}\sum_{|{\boldsymbol R}|=k}
\prod_{i=1}^N \prod_{s \in R_i}
\frac{(-1)^{i-1}\,e^{\iu \frac{N-2}2 E_{i\varnothing}(s)}}{\prod_{j=1}^N(2\iu)\sin\frac12 E_{ij}(s)}.
\ee

As emphasized, our viewpoint is that the real topological string computation defines the rules to describe the properties of the orientifold surface operator in the $\SU(N)$ gauge theory. Let us now discuss the effect of the additional signs from the perspective of the gauge theory, to gain insight into the additional ingredients in $\mathcal O_\Omega$ beyond the $\epsilon_2$-shift. The orientifold is acting with different (alternating) signs on the different Young diagram degrees of freedom associated to the edges in the web diagram. This might imply an action with different signs on the states charged under the corresponding 
Cartans (alternating when ordered as determined by the 5d real Coulomb branch parameters). It would be interesting to gain a more direct gauge theory insight into the definition of this orientifold action.

Before concluding, we would like to mention an important point. We have used  the $\epsilon_2$-shift exactly as in the conifold case in \cref{sec:invariant}, and obtained  a result very close to the real topological vertex computation. However, one should keep in mind that the BPS states of the $\SU(N)$ theory have a much richer structure. Therefore, the additional signs are of crucial importance to reproduce the correct results for the complete orientifold action on the theory.

For instance, if we isolate the unoriented contribution from \cref{signed-fmla}, the extra signs are crucial to produce certain non-zero real BPS multiplicities. This can be checked explicitly e.g.\ for $\SU(2)$ using the results from \cref{sec:enum}. For instance, consider the real BPS multiplicities $n_{d_1,d_2}^g$ for M2-branes wrapped with degrees $d_1$, $d_2$ on the homology classes $B$ and $F$, respectively.
Already at $g=0$ we have $n_{1,0}^0=-2$ but $n_{0,1}^0=0$. The extra signs are crucial to produce a non-zero result for the unoriented contribution of the vector multiplet from the M2-brane on $B$ (which using \cref{SUN.gauge.real} would give zero contribution). Similar considerations can be drawn for many others of the enumerative results in \cref{sec:enum}.

\section{Conclusions}
\label{conclusions}

In this work we have explored the extension of the correspondence between topological strings on toric CY three-folds and 4d/5d supersymmetric gauge theories with 8 supercharges to systems with orientifolds with real codimension 2 fixed locus. On the topological string side, we have focused on quotients which produce the real topological string of \cite{Walcher:2007qp}, because of its remarkably simple physical realization in M-theory. We have analyzed the properties of the systems, and emphasized their behavior under flops of the geometry.

The real topological string amplitudes define the properties of a new kind of surface defect in the corresponding gauge theory. We have rephrased the amplitudes in a form adapted to a gauge theory interpretation,
by means of a newly defined twisted Nekrasov partition function,
and we have taken the first steps towards providing an intrinsically gauge-theoretic interpretation of the twisting operator.

It would be interesting to complete the gauge theory interpretation of the twist operator. The partition function obtained by the simple $\epsilon_2$ shift does not distinguish invariant states from non-invariant states. Therefore, M2-branes wrapping $F$ and M2-branes wrapping $B$ essentially give the same contribution to the partition function \cref{SUN.gauge.real}. However, in general, they would give a different contribution in the real topological string amplitude since the former correspond to non-invariant states and the latter correspond to invariant states. Hence, another implementation may be related to some operation that distinguishes the invariant states from the non-invariant states. There may be also a possibility to shift the Coulomb branch moduli like the ordinary orbifolded instantons\cite{Kanno:2011fw}.

A way to complete the gauge theory interpretation may be to give a more specific description of the orientifold in the ADHM quantum mechanics. In \cite{Kanno:2011fw}, instantons with a surface defect were identified with orbifolded instantons via a chain of dualities of string theory. It would be interesting to extend their reasoning to our case and determine an effect of the orientifold defect in the ADHM quantum mechanics. Once we identify the effect in the ADHM quantum mechanics, then we may proceed in the standard localization technique with it.

There are several other interesting directions worth exploring:
\begin{itemize}

\item It would be interesting to exploit the M-theory picture to develop a refined real topological vertex formalism, and to compare it with the gauge theory computations.

\item As explained, there are different kinds of O4-planes in the physical type IIA picture,
which correspond to different M-theory lifts, and different unoriented topological strings (albeit, with intricate relations).
We hope to explore the gauge theory description of those in future works.

\item It would also be interesting to consider the addition of extra D4-branes,
either on top of the O4-plane or possibly on other lagrangian 3-cycles,
to describe the unoriented version of the relation of open topological strings
and vortex counting on surface defects\cite{Dimofte:2010tz,Bonelli:2011fq}.

\end{itemize}

We hope to come back to these question in the future.

\subsection*{Acknowledgments}

We thank
G.~Bonelli, O.~Foda, H-C.~Kim, D.~Krefl, K.S.~Narain, N.~Nekrasov, S.~Pasquetti, Y.~Tachikawa, A.~Tanzini, J.~Walcher, J.F.~Wu, M.~Young and J.~Zhou
for useful discussions. We also thank the organizers of the ``V Workshop on Geometric Correspondences of Gauge Theories'', at SISSA, where part of this work was presented and developed. HH and AU are partially supported by the grants FPA2012-32828 from the MINECO, the ERC Advanced Grant SPLE under contract ERC-2012-ADG-20120216-320421 and the grant SEV-2012-0249 of the ``Centro de Excelencia Severo Ochoa" Programme. The work of HH is supported also by the REA grant agreement PCIG10-GA-2011-304023 from the People Programme of FP7 (Marie Curie Action) and also in part by Perimeter Institute for Theoretical Physics where a part of the work was done. Research at Perimeter Institute is supported by the Government of Canada through Industry Canada and by the Province of Ontario through Ministry of Economic Development \& Innovation.
The work of NP is partially supported by
the COST Action MP1210 ``The string Theory Universe''
under ECOST-STSM-MP1210-120115-051514
and by the Italian National Group of Mathematical Physics (GNFM-INdAM).

\appendix

\section{Real topological string}
\label{app:background}

\subsection{M-theory interpretation of the real topological string}
\label{sec:m-theory}

\paragraph{BPS state counting}

As already observed, we consider tadpole canceling configurations,
such that the M-theory lift of the O4/D4 system is smooth, i.e.~there are no fixed points.
This guarantees that locally, before moving around the M-circle, the physics looks like in the oriented case.
The $\SU(2)_L \times \SU(2)_R$ group is broken by the orientifold to its Cartan generators,
which are enough to assign multiplicities to the 5d BPS states.

There will be two kinds of states, as follows.
First, those not invariant under the orientifold,
will have their orientifold image curve somewhere in the covering $\IX$,
and will contribute to closed oriented amplitudes (thus, we can neglect them).
Second, those corresponding to curves mapped to themselves by the involution;
their overall $\IZ_2$ parity is determined by their $\SU(2)_L$ and $\SU(2)_R$ multiplet structure:
as in the original GV case, the $R$ part parametrizes our ignorance about cohomology of D-brane moduli space,
while the $L$ part is cohomology of Jacobian, and we know how to break it explicitly,
due to our simple orientifold action.
Let us split the second class of states, for fixed genus $\hg$ and homology class $\beta$, according to their overall parity
\be
\gv'_{\hg,\beta} = \gvp + \gvm.
\ee

Once we move on the circle, these invariant states acquire integer or half-integer momentum, according to their overall parity;
this is taken into account in the 2d Schwinger computation, that finally yields,
after removing even wrapping states that correspond again to closed oriented sector, the numbers
\be
\hgv := \gvp - \gvm
\ee
that appear in \cref{final-trace}.
The detailed computation is described in\cite{unoriented-gv}.

\paragraph{Tadpole cancellation}

The requirement that physical tadpoles are canceled has an interesting implication for real topological amplitudes,
for geometries with $H_1(L;\IZ)=\IZ_2$ and $H_2(\IX,L;\IZ)=\IZ$;
these include well-studied examples like the real quintic or real local $\CP^2$.
We discuss this for completeness, even though the geometries in the main text do not have this torsion homology on $L$.

Denote the degree of a map by $d \in H_2(\IX,L;\IZ)$.
By looking at the appropriate exact sequence in homology, one can see that crosscaps contribute an even factor to $d$,
while boundaries may contribute even or odd factor.
Moreover, by looking at the M-theory background form $C_3$,
one can see that it contributes to the central charge a factor of $\iu /2$ for every crosscap, i.e.~$\RP^2$,
that surrounds the O4-plane.
This translates into a minus sign once Poisson resummation is performed,
more precisely a $(-1)^{mc}$ sign, where $c$ is the crosscap number and $m$ wrapping number around the circle
(recall the states we are interested in have odd $m$.)

Finally, since boundaries do not receive such contributions and one can show,
with a heuristic argument regarding real codimension one boundaries in the moduli space of stable maps,
that there is a bijection between curves that agree except for a replacement of a crosscap with a (necessarily even-degree) boundary,
we conclude that these two classes of curves cancel against each other.
This implies that the only contributions may arise from odd-degree-boundary curves,
and it is written as a restriction $\chi=d \bmod 2$ on the summation in \cref{final-trace},
as it has been proposed in \cite{Walcher:2007qp} based on the fact that these two contributions to the topological amplitude are not mathematically well-defined separately,
and the above mentioned prescription produces integer BPS multiplicities.

\subsection{Real topological vertex}
\label{sec:real-vertex}

The real topological vertex\cite{Krefl:2009mw} is a technique that allows to compute the all genus topological string partition functions,
in the presence of toric orientifolds, namely a symmetry of the toric diagram with respect to which we quotient.
This corresponds to an involution $\sigma$ of $\IX$, and it introduces boundaries and crosscaps in the topological string theory.
We restrict to unrefined quantities, since at the moment real topological vertex technology is only available for that setup.

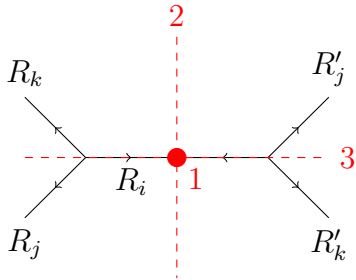
\begin{figure}[htb]
\centering
\begin{tikzpicture}[node distance=2cm, auto,scale=.4]
\draw[->-] (-3,0) to (0,0);
\draw[->-] (3,0) to (0,0);
\draw[->-] (3,0) to (5,2);
\draw[->-] (3,0) to (5,-2);
\draw[->-] (-3,0) to (-5,2);
\draw[->-] (-3,0) to (-5,-2);

\node[anchor=south] at (5,2) {$R'_j$};
\node[anchor=north] at (5,-2) {$R'_k$};
\node[anchor=north] at (-1.5,0) {$R_i$};
\node[anchor=north] at (-5,-2) {$R_j$};
\node[anchor=south] at (-5,2) {$R_k$};

\color{red}

\draw[dashed] (0,4) to (0,-4);
\node[anchor=south] at (0,4) {2};
\draw[dashed] (-5,0) to (5,0);
\node[anchor=west] at (5,0) {3};

\draw[fill] (0,0) circle (.3);
\node[anchor=north west] at (0,0) {1};
\end{tikzpicture}
\caption{Three involutions for a generic internal leg; notice that each involution requires a specific symmetry to be present.}
\label{fig:top-vert-inv}
\end{figure}

The recipe morally amounts to taking a square root of the topological vertex amplitude
of the corresponding oriented parent theory, as follows.
First, we observe that contributions from legs and vertices that are not fixed by the involution
can be dealt with using standard vertex rules,
and they automatically give rise to a perfect square once paired with their image.
We then only need to explain how to deal with a fixed edge connecting two vertices.
Their contribution to the partition function is given by a factor
\be
\label{contrib-fixed}
\sum_{R_i} C_{R_j R_k R_i}(- e^{-t_i})^{|R_i|} (-1)^{n|R_i|} q^{\frac12 n \left(||R_i||^2-||R_i^t||^2\right)}
C_{R'_j R'_k R^t_i},
\ee
where $C :=C(q,q)$ was introduced in \cref{vertex-functions},
and notation corresponds to \cref{fig:top-vert-inv}.
Here $n:=\det(v_{j'}, v_j)$,
where $v_m$ represents an outgoing vector associated to leg $m$.

There are three cases: the involution can act as a point reflection at the center of the line (1),
a reflection at the line perpendicular to the compact leg (2),
or a reflection along the compact leg (3).
The case interesting for us is (2),
namely a leg that is not point-wise fixed by the involution,
and where the representations in one vertex are mapped to representations in the other.
In this case, no restriction is imposed on the internal representation $R_i$,
while leg $j$ is mapped to leg $k'$ and similarly $k \to j'$.
This imposes $R_j=R_k'^t$ and $R_k=R_j'^t$,
where the transposition is implemented
since the involution introduces an orientation-reversal of the plane in \cref{fig:top-vert-inv}.
By exploiting the symmetry of function $C$
\be
\label{permutation}
C_{AB^tC}=q^{\frac{||A||^2-||A^t||^2+||B^t||^2-||B||^2+||C||^2-||C^t||^2}2}
C_{BA^tC^t},
\ee
we can rewrite \cref{contrib-fixed} as a perfect square, and take the square root:
\be
\label{real.top.result}
\sum_{R_i} C_{R_j R_k R_i} e^{-\frac12 t_i |R_i|} (-1)^{\frac12 (n+1) s(R_i)}
q^{\frac14 (n-1) (||R_i||^2-||R_i^t||^2) + \frac14 (||R_k^t||^2-||R_k||^2)+\frac14 (||R_j||^2-||R_j^t||^2)}.
\ee
We introduced a sign in \cref{real.top.result}
\be
\label{sign.real}
(-1)^{\frac12 (n+1) s(R_i)},
\ee
determined by $s(R)= |R| \pm c(R)$,
where $c(R)$ is defined via $|R|-c(R)=2\sum_i R(2i)$.
Finally, there is a global prescription for the choice of $c(R_i)$ vs.\ $c(R_i^t)$.

\section{Enumerative checks}
\label{sec:enum}

We compute the real GV invariants of the $\SU(N)$ geometry with the involution considered in \cref{sec:sun-unoriented}. We describe some numerical checks that the topological vertex amplitudes indeed produce integer BPS multiplicities,
corresponding to the proposed BPS state counting for the real topological string\cite{Walcher:2007qp,unoriented-gv}.
The enumerative checks support the new result of the real topological string partition function for $\SU(N)$ geometry in \cref{sec:sun-unoriented}.

After removing the purely closed oriented contribution, we perform an expansion
\be
\label{unoriented}
Z^\text{unor}=
\frac{Z^\text{real}}{\sqrt{Z^\text{or}}} =
1 + Z^\text{real}_\text{1-inst} u^\frac12 +
\left(Z^\text{real}_\text{2-inst} - \frac12 Z_\text{1-inst}\right) u
+ \mathcal O \left( u^\frac32 \right).
\ee
Note that the perturbative contribution does not contribute to the unoriented string part in the current choice of involution.
From \cref{unoriented}, we can compute the real GV invariants:
they are the numbers $n$ appearing if we rewrite it using \cref{final-trace},
\be
\label{GV}
Z^\text{unor} = \exp
\sum_{d_1, d_2, g;\text{ odd }k} \frac{n_{d_1, d_2}^g}k \left(2\iu\sin\frac{k\epsilon}2\right)^{g-1}
Q_B^{\frac{kd_1}2} Q_F^{\frac{kd_2}2},
\ee
where we focus on the $\SU(2)$ computation;
from \cref{vertex-su-n} with $N=2$ we obtain
\be
Z^\text{real}_\text{1-inst}=\frac1{2\sin\frac{\epsilon}2 \sin t}
\ee
and
\be
Z^\text{unor}_{u^1}=-\frac38 \frac1{\sin^2\frac{\epsilon}2\left(\cos 2t - \cos\epsilon\right)} -
\frac1{16}\frac1{\sin^2 t \sin^2\frac{\epsilon}2}.
\ee
For illustration, we obtain real GV numbers up to $d_2\leq 6$ and $g\leq 6$.
We have $n_{1, d_2}^0 = -2$, for $d_2=0, 2, 4, 6$, and the others are zero.
For $n_{2, d_2}^g$, we have
\be
\begin{tabu}{c|ccccccc}
d_2 \setminus g & 0 & 1 &  2 & 3 & 4 & 5 & 6\\
\hline
0 & 0 & 0 & 0 & 0 & 0 & 0 & 0 \\
1 & 0 & 0 & 0 & 0 & 0 & 0 & 0\\
2 & 0 & 3& 0 & 0 & 0 & 0 & 0\\
3 & 0 & 0 & 0 & 0 & 0& 0 & 0\\
4 & 0 & 12 & 0 & 3 & 0 & 0 & 0\\
5 & 0 & 0 & 0 & 0 & 0 & 0 & 0\\
6 & 0 & 30 & 0 & 18 & 0 & 0 & 0
\end{tabu}
\ee
For $d_1=3$ we have
\be
\begin{tabu}{c|ccccccc}
d_2 \setminus g & 0 & 1 &  2 & 3 &  4 & 5 & 6\\
\hline
0 & 0 & 0 & 0 & 0 & 0 & 0 & 0 \\
1 & 0 & 0 & 0 & 0 & 0 & 0 & 0\\
2 & -6 & 0& -4 & 0 & 0 & 0 & 0\\
3 & 0 & 0 & 0 & 0 & 0& 0 & 0\\
4 & -28 & 0 & -58 & 0 & -28 & 0 & -4\\
5 & 0 & 0 & 0 & 0 & 0 & 0  & 0\\
6 & -82 & 0 & -324 & 0 & -362 & 0  &  -184
\end{tabu}
\ee
For $d_1=4$ we have
\be
\begin{tabu}{c|ccccccc}
d_2 \setminus g & 0 & 1 &  2 & 3 &  4 & 5 & 6\\
\hline
0 & 0 & 0 & 0 & 0 & 0 & 0 & 0 \\
1 & 0 & 0 & 0 & 0 & 0 & 0 & 0\\
2 & 0 & 12& 0 & 5 & 0 & 0 & 0\\
3 & 0 & 0 & 0 & 0 & 0& 0 & 0\\
4 & 0 & 153 &  0 & 268 & 0 & 177 & 0\\
5 & 0 & 0 & 0 & 0 & 0 & 0 & 0\\
6 & 0 & 900 & 0 & 3107 & 0 & 4670 & 0
\end{tabu}
\ee
For $d_1 = 5$ we have
\be
\begin{tabu}{c|ccccccc}
d_2 \setminus g & 0 & 1 &  2 & 3 &  4 & 5 & 6\\
\hline
0 & 0 & 0 & 0 & 0 & 0 & 0 & 0 \\
1 & 0 & 0 & 0 & 0 & 0 & 0 & 0\\
2 & -12 & 0& -20 & 0 & -6 & 0 & 0\\
3 & 0 & 0 & 0 & 0 & 0& 0 & 0\\
4 & -156 & 0 & -744 & 0 & -1212 & 0 & -962\\
5 & 0 & 0 & 0 & 0 & 0 & 0  & 0\\
6 & -990 & 0 & -8518 & 0 & -27704 & 0  &  -49814
\end{tabu}
\ee
For $d_1=6$ we have
\be
\begin{tabu}{c|ccccccc}
d_2 \setminus g & 0 & 1 &  2 & 3 &  4 & 5 & 6\\
\hline
0 & 0 & 0 & 0 & 0 & 0 & 0 & 0 \\
1 & 0 & 0 & 0 & 0 & 0 & 0 & 0\\
2 & 0 & 30& 0 & 30 & 0 & 7 & 0\\
3 & 0 & 0 & 0 & 0 & 0& 0 & 0\\
4 & 0 & 900 &  0 & 3293 & 0 & 5378 & 0\\
5 & 0 & 0 & 0 & 0 & 0 & 0 & 0\\
6 & 0 & 10255 & 0 & 70128  & 0 & 232826  & 0
\end{tabu}
\ee

\paragraph{Bound on genus}

One possible check we can perform is obtain the maximal $g$ for given $d$\cite{Katz:1999xq}.
To do this, let us take diagonal combinations
$n^g_d:= \sum_{d_1+d_2=d} n_{d_1,d_2}^g$:
\be
\label{GVtop}
\begin{tabu}{c|cccc}
d \setminus g & 0 & 1 & 2 & 3\\
\hline
1 & -2 & 0 & 0 & 0\\
2 & 0 & 0 & 0 & 0\\
3 & -2 & 0 & 0 & 0\\
4 & 0 & 3 & 0 & 0\\
5 & -8 & 0 & -4 & 0\\
6 & 0 & 24 & 0 & 8 
\end{tabu}
\ee
They satisfy tadpole cancellation $d=\chi=g-1 \mod 2$.
For fixed $d=d_1+d_2$, a smooth curve in $\CP^1 \times \CP^1$ has genus $g=(d_1-1)(d_2-1)$, which is the top genus.
For $d=1$, we have a non-zero contribution from $d_1 = 1$ and $d_2=0$ for the involution we are considering.
The top genus contribution appears from $g=0$.
The Gopakumar-Vafa invariant is then related to the Euler characteristic\footnote
{Recall that $\eu{\CP^m}=m+1$ and $\eu{\RP^m}=1,0$ for $m$ even/odd respectively.}
of the moduli space\cite{Katz:1999xq}
\be
n_1^0 = n_{1,0}^0 = - \eu{\CP^1} = -2,
\ee 
which is consistent with \cref{GVtop}.

The dimension of the moduli space of a curve inside $\CP^1 \times \CP^1$ may be understood as follows.
The Gopakumar-Vafa invariants $n_{d_1,d_2}^g$ are related to M2-branes wrapping a two-cycle
\be
d_1\left[\CP^1_B\right] + d_2\left[\CP^1_F\right],
\ee
where $\left[\CP^1\right]$ represents a divisor class of $\CP^1$.
The degrees are also related to the degree of a polynomial that represents the curve by
\be
\sum_{a_1,a_2,b_1,b_2} \alpha_{a_1,a_2,b_1,b_2} X_0^{a_1}X_1^{a_2} \tilde X_0^{b_1} \tilde X_1^{b_2} = 0,
\qquad a_1+a_2 = d_2, \quad b_1+b_2=d_1,
\ee
where $(X_0, X_1)$ are homogeneous coordinates of $\CP^1_B$
and $(\tilde X_0, \tilde X_1)$ are homogeneous coordinates of $\CP^1_F$.
We are now considering the case with $d_1=1$, $d_2 = 0$, which is
\be
\label{degree1}
\alpha_{0,0,1,0} \tilde X_0 + \alpha_{0,0,0,1} \tilde X_1 = 0.
\ee
$\alpha_{0,0,1,0}$ and $\alpha_{0,0,0,1}$ still take value in $\IC$ and \cref{degree1} is still a complex equation.
Therefore, the moduli space parametrized by $\alpha_{0,0,1,0}$ and $\alpha_{0,0,0,1}$ is $\CP^1$.
On the other hand, the deformation space of the curve class $d_2\left[\CP^1_F\right]$
gives a real projective space due to the involution acing on $\CP^1_B$.
In general, the moduli space may be given by $\CP^{d_1 } \times \RP^{d_2}$.

We can also consider the case $d=2$.
Then the top genus comes from $g=0$.
But this contribution will be absent since this does not satisfy the tadpole condition $d = g-1 \mod 2$.
This is also consistent with \cref{GVtop}.

The degree 3 case is also the same, namely
\be
n_3^0 = n_{1,2}^0 = -\eu{\CP^1} \eu{\RP^2} = -2.
\ee 
One may do similarly for $d=4$.
The maximal genus is 1 and hence 3 in \cref{GVtop} will be related to the Euler characteristic of the moduli space: we find
\be
n_4^1 = n_{2,2}^1 = (-1)^2 \eu{\CP^2} \eu{\RP^2} = 3.
\ee
For $d=5$ we find 
\be
n_5^2 = n_{3,2}^2 = (-1)^3 \eu{\CP^3} \eu{\RP^2} = -4,
\ee
which is again consistent with the obtained result.
For $d=6$, the top genus is $g=4$.
But this does not satisfy the tadpole condition and hence $n_6^4 = 0$.
That is also consistent with the result.

\section{Definitions and useful identities}
\label{sec:help}

In this appendix we list some useful identities for quantities appearing in the topological vertex amplitudes.
For non-trivial ones, we cite a reference where a proof can be found.

One can prove\cite[eqs.~(2.9) and (2.16)]{Awata:2008ed} the identity
\be
\label{symmetric}
\prod_{i,j=1}^\infty \left( 1-Qq^{i+j-1-R_1(j)-R_2(i)} \right)=\\
\prod_{i,j=1}^\infty \left( 1-Qq^{i+j-1} \right)
\prod_{s \in R_2} \left( 1-Qq^{-a_{R_1^t}(s) -1 -l_{R_2}(s)} \right)
\prod_{s \in R^t_1} \left( 1-Qq^{a_{R_2}(s) +1 +l_{R^t_1}(s)} \right).
\ee
From definition $||R||^2:= \sum_i R(i)^2$, some simple combinatorial identities follow
\be
\sum_i i \nu(i)  = \frac{||\nu^t||^2}{2} + \frac{|\nu|}{2},
\qquad
\sum_{(i,j) \in \nu} \tilde \nu^t (j) = \sum_{(i,j) \in \tilde \nu} \nu^t(j),
\qquad
\sum_{s \in \nu} l_{\nu}(s) = \frac{||\nu||^2}{2} - \frac{|\nu|}{2},
\ee
which imply
\be
\sum_{(m,n) \in R_i} a_{R_j} (m,n) =
\frac12 \left( ||R^t_j||^2+|R_j|-||R^t_i||^2-|R_i|\right)
+ \sum_{(m,n) \in R_j} a_{R_i} (m,n).
\ee
Using the above we get
\be
\label{combinatorics}
\prod_{s \in R_i} \left( 1-e^{\iu E_{ij}(s)} \right)^{-1}
\prod_{s \in R_j} \left( 1-e^{-\iu E_{ji}(s)} \right)^{-1}
=\\
(-1)^{|R_i|} e^{\iu \frac{t_{ji}}2 (|R_i|+ |R_j|)}
e^{\iu \frac{\epsilon}4 [||R_j^t||^2-||R_j||^2-||R_i^t||^2+||R_i||^2]}
\prod_{s \in R_i} \frac1{2\iu\sin \frac{E_{ij}(s)}2}
\prod_{s \in R_j} \frac1{2\iu\sin \frac{E_{ji}(s)}2}.
\ee

Finally, the important identity for skew-Schur functions\cite[page 93]{macdonald-symm-func}
\be
\label{schur-id}
\sum_\mu s_{\mu/\eta}(x) s_{\mu/\xi}(y) =
\prod_{i,j=1}^\infty (1-x_iy_j)^{-1} \sum_\mu s_{\xi/\mu}(x) s_{\eta/\mu}(y)
\ee
gives upon iteration\cite[Lemma 3.1]{zhou-curve}
\be
\sum_{\substack {\nu^1, \dots, \nu^N \\ \eta^1, \dots, \eta^{N-1}}}
\prod_{k=1}^N s_{\nu^k/\eta^{k-1}}(x^k) Q_k^{|\nu^k|} s_{\nu^k/\eta^k}(y^k)
= \prod_{\substack {1 \leq k < l \leq N+1 \\ i, j \geq 1}}
\left( 1- Q_kQ_{k+1} \cdots Q_{l-1} x^k_iy^{l-1}_j \right)^{-1},
\ee
where $\eta^0=\varnothing=\eta^N$ and $x^k=(x^k_1, x^k_2, \dots)$.

\bibliographystyle{JHEP}
\bibliography{biblio}
\end{document}